\documentclass[10pt]{article}

\usepackage{amsmath}
\usepackage{amsfonts}
\usepackage{amssymb}
\usepackage{natbib}
\usepackage{graphicx}
\usepackage{bm}

\usepackage[margin=1.2in]{geometry}

\usepackage{times}

\usepackage{setspace}
\def\listofnotations{
\begin{tabbing}
$$~~~~~\=\parbox{5in}{ }\\
\addnotation \alpha: {Biot's coefficient, (Eqn. \ref{alpha2})}{symbol:aplha}
\addnotation \beta: {compressibility of the fluid phase}{symbol:beta}
\addnotation \gamma_w: {specific weight of the fluid phase}{symbol:gamma_w}
\addnotation \delta_{ij}: {Kronecker delta}{symbol:delta_ij}
\addnotation \epsilon_{ij}: { components of  strain tensor}{symbol:epsilon_ij}
\addnotation \bm{\epsilon}: {strain vector, Eqn. (\ref{sigma_voigt})}{symbol:epsilon_v}
\addnotation \epsilon: {volumetric strain, $\epsilon = \epsilon_{xx} +\epsilon_{yy}+\epsilon_{zz}$ }{symbol:epsilon}
\addnotation \zeta: {increment of fluid content, (Eqns. \ref{increment_fl_1}, \ref{zeta_1})}{symbol:zeta}
\addnotation \eta: {poroelastic stress coefficient, Eqn. \ref{por_stress_coeff}}{symbol:eta}
\addnotation \lambda: {Lam\'{e}'s coefficient, (Eqn. \ref{shear_modulus})}{symbol:lambda}
\addnotation \lambda_u: {undrained Lam\'{e}'s coefficient, (Eqns. \ref{lame_undrained}, \ref{Lame_undrained_2} ,
\ref{Lame_undrained_3}, \ref{poisson_undrained_4}, \ref{undrained_shear_modulus}) }{symbol:lambda_u}
\addnotation \nu: {Poisson's ratio}{symbol:nu}
\addnotation \nu_u: {undrained Poisson's ratio, (Eqns. \ref{poisson_undrained}, \ref{poisson_undrained_3}, \ref{poisson_undrained_4})}{symbol:nu_u}
\addnotation \rho_f: {fluid density}{symbol:rho_f}
\addnotation \sigma: {mean stress, $\sigma = (\sigma_{xx}+\sigma_{yy}+\sigma_{zz})/3 $}{symbol:sigma}
\addnotation \sigma_{ij}: {components of stress tensor}{symbol:sigma_ij}
\addnotation \bm{\sigma}: {stess vector, Eqn. (\ref{sigma_voigt})}{symbol:sigma_v}
\addnotation \sigma': {mean effective stress}{symbol:sigma_eff}
\addnotation \sigma'_{ij}: {components of effective stress tensor}{symbol:sigma_eff_ij}

\addnotation : {}{symbol:}
\addnotation b_i: {i-th component of the body forces}{symbol:b_i}
\addnotation c: {coefficient of consolidation, (Eqn. \ref{coefficient_of_consolidation_3d_1}, 
  \ref{coefficient_of_consolidation_3d_1_2},\ref{coefficient_of_consolidation_3d}, \ref{coefficient_of_consolidation_3d_2}, 
  \ref{coefficient_of_consolidation_3d_3}, \ref{coefficient_of_consolidation_3d_4} )}{symbol:c}
\addnotation k: {coefficient of permeability}{symbol:k}
\addnotation m_v:  {coefficient of one dimensional compressibility under drained conditions, (Eqns. \ref{constitutive_law}, \ref{volume_change1})}{symbol:mv}
\addnotation \bold m: {vector form of Kronecker's delta, Eqn. (\ref{m_voigt}) }{symbol:m_v}
\addnotation n: {porosity}{symbol:n}
\addnotation p: {pore pressure, (Eqn. \ref{zeta_press})}{symbol:p}
\addnotation \bold q: {specific discharge, (Eqn. \ref{discharge}) }{symbol:q}
\addnotation u_i: {i-th component of the displacements vector}{symbol:u_i}
\addnotation \bold v_f: {velocity of the fluid phase}{symbol:v_f}
\addnotation \bold v_s: {velocity of the solid phase}{symbol:v_s}

\addnotation : {}{symbol:}
\addnotation B: {Skempton's pore pressure coefficient, (Eqns. \ref{Bishop_B}, \ref{B_2})}{symbol:B}
\addnotation B': {pore pressure coefficient under uniaxial strain (or loading efficiency), (Eqns.  \ref{B_uniaxial}, \ref{B_uniaxial_3}, \ref{B_uniaxial_undrained},\ref{B_uniaxial2}) }{symbol:B'}
\addnotation C: {compressibility of the solid skeleton (drained compressibility of the porous medium) }{symbol:C}
\addnotation C_s: {compressibility of the solid phase (considered equal to the compressibility of solid grains), (Eqn.  \ref{compress_solidPhase})}{symbol:Cs}
\addnotation C_{\phi}: {unjacketed pore compressibility, (Eqn.  \ref{compress_pore})}{symbol:Cs}
\addnotation D: {elasticity stiffness matrix, Eqn. (\ref{D_matrix_3d})}{symbol:D}
\addnotation E: {Young's modulus}{symbol:E}
\addnotation E_u: {undrained Young's modulus, (Eqn. \ref{undrained_E})}{symbol:Eu}
\addnotation G: {shear modulus, (Eqn. \ref{shear_modulus})}{symbol:G}
\addnotation K: {bulk modulus, (Eqn. \ref{shear_modulus}) }{symbol:K}
\addnotation K_u: {undrained bulk modulus, (Eqns. \ref{k_undrained}, \ref{k_undrained_gassman}, \ref{undrained_shear_modulus}) }{symbol:Ku}
\addnotation K_f: {bulk modulus of the fluid phase }{symbol:Kf}
\addnotation K_s: {bulk modulus of the solid phase }{symbol:Ks}
\addnotation M: {Biot's modulus, (Eqn.  \ref{biot_modulus}, \ref{biot_M_ku})}{symbol:M}
\addnotation S: {uniaxial storage coefficient, (Eqns.  \ref{uniaxial_storage}, \ref{uni_storage_def},
                  \ref{coefficient_of_consolidation_3d}, \ref{uniaxial_storage_2})}{symbol:S}
\addnotation S_{\epsilon}: {storage coefficient, (Eqns. \ref{storativity_Cvoid}, \ref{storativity},  \ref{storativity_3}) }{symbol:Se}
\addnotation S_r: {degree of saturation}{symbol:Sr}

\addnotation : {}{symbol:}
\addnotation \nabla_s: {symmetric gradient operator, Eqn. (\ref{symm_grad_3d}) }{symbol:nabla_symmetric}

\end{tabbing}
 \clearpage}
\def\addnotation #1: #2#3{$#1$\> \parbox{5in}{#2}\\}

\makeatletter
\def\cleardoublepage{\clearpage\if@twoside \ifodd\c@page\else
    \hbox{}\thispagestyle{empty}\newpage\if@twocolumn\hbox{}\newpage\fi\fi\fi}
\makeatother

\usepackage{color}
\definecolor{gray}{cmyk}{0,0,0,0.28}

\begin{document}

\title{\huge An introduction to linear poroelasticity}

\maketitle
\begin{center}
\author{ \large Andi Merxhani \footnote{PhD, Jesus College, University of Cambridge.}
                         \\  \normalsize am3232@caa.columbia.edu}
\end{center}

\begin{center}
\date { \today}
\end{center}

\thispagestyle{empty}

\section*{}
This study is an introduction to the theory of poroelasticity expressed in terms of Biot's theory 
of three-dimensional consolidation. 
The point of departure in the description are the basic equations of elasticity (i.e. constitutive law, equations of equilibrium 
in terms of stresses, and the definition of strain), together with the principle of effective stress, and Darcy's 
law of fluid flow in porous media. These equations, together with the principle of mass conservation, are the only premises 
used to derive Verruijt's formulation of poroelasticity as used in soil mechanics. The equation of fluid mass balance derived
in this work is an extension to Verruijt's original derivation, since it also considers the effect of the unjacketed 
pore compressibility (i.e. it accounts for solid phase not being composed of a single constituent, 
and for the existence of occluded voids and/or cracks within the solid skeleton.)
Verruijt's formulation uses a drained description where pore pressure is an independent variable. 
Next, the increment of fluid content is defined and its constitutive law is derived - with its derivation following
naturally from the equation of fluid mass balance. 
Pore pressure, storage, and undrained poroelastic coefficients are also introduced and useful relations are 
proven. Where appropriate, the physical meaning of these coefficients is proven mathematically.
Equations of equilibrium and fluid mass conservation are subsequently expressed in terms of the increment of fluid content 
and undrained coefficients, leading to an undrained description of poroelasticity. Thus Verruijt's approach 
is extended to Rice and Cleary's formalism. 
This approach to 
poroelasticity is useful for its simplicity. It does not require the ad-hoc definition of poroelastic constants
and that of an elastic energy potential. Instead, it is a direct extension to isothermal linear elasticity that 
accounts for the coupling of skeletal deformations and fluid behaviour.

\newpage

\tableofcontents




\section*{List of notations}


\begin{tabbing}
$$~~~~~\=\parbox{5in}{ }\\
\addnotation \alpha: {Biot's coefficient, (Eqn. \ref{alpha2})}{symbol:aplha}
\addnotation \beta: {compressibility of the fluid phase}{symbol:beta}
\addnotation \gamma_w: {specific weight of the fluid phase}{symbol:gamma_w}
\addnotation \delta_{ij}: {Kronecker delta}{symbol:delta_ij}
\addnotation \epsilon_{ij}: { components of  strain tensor}{symbol:epsilon_ij}
\addnotation \bm{\epsilon}: {strain vector, Eqn. (\ref{sigma_voigt})}{symbol:epsilon_v}
\addnotation \epsilon: {volumetric strain, $\epsilon = \epsilon_{xx} +\epsilon_{yy}+\epsilon_{zz}$ }{symbol:epsilon}
\addnotation \zeta: {increment of fluid content, (Eqns. \ref{increment_fl_1}, \ref{zeta_1})}{symbol:zeta}
\addnotation \eta: {poroelastic stress coefficient, Eqn. \ref{por_stress_coeff}}{symbol:eta}
\addnotation \lambda: {Lam\'{e}'s coefficient, (Eqn. \ref{shear_modulus})}{symbol:lambda}
\addnotation \lambda_u: {undrained Lam\'{e}'s coefficient, (Eqns. \ref{lame_undrained}, \ref{Lame_undrained_2} ,
\ref{Lame_undrained_3}, \ref{poisson_undrained_4}, \ref{undrained_shear_modulus}) }{symbol:lambda_u}
\addnotation \nu: {Poisson's ratio}{symbol:nu}
\addnotation \nu_u: {undrained Poisson's ratio, (Eqns. \ref{poisson_undrained}, \ref{poisson_undrained_3}, \ref{poisson_undrained_4})}{symbol:nu_u}
\addnotation \rho_f: {fluid density}{symbol:rho_f}
\addnotation \sigma: {mean stress, $\sigma = (\sigma_{xx}+\sigma_{yy}+\sigma_{zz})/3 $}{symbol:sigma}
\addnotation \sigma_{ij}: {components of stress tensor}{symbol:sigma_ij}
\addnotation \bm{\sigma}: {stess vector, Eqn. (\ref{sigma_voigt})}{symbol:sigma_v}
\addnotation \sigma': {mean effective stress}{symbol:sigma_eff}
\addnotation \sigma'_{ij}: {components of effective stress tensor}{symbol:sigma_eff_ij}

\addnotation : {}{symbol:}
\addnotation b_i: {i-th component of the body forces}{symbol:b_i}
\addnotation c: {coefficient of consolidation, (Eqn. \ref{coefficient_of_consolidation_3d_1}, 
  \ref{coefficient_of_consolidation_3d_1_2},\ref{coefficient_of_consolidation_3d}, \ref{coefficient_of_consolidation_3d_2}, 
  \ref{coefficient_of_consolidation_3d_3}, \ref{coefficient_of_consolidation_3d_4} )}{symbol:c}
\addnotation k: {coefficient of permeability}{symbol:k}
\addnotation m_v:  {coefficient of one dimensional compressibility under drained conditions, (Eqns. \ref{constitutive_law}, \ref{volume_change1})}{symbol:mv}
\addnotation \bold m: {vector form of Kronecker's delta, Eqn. (\ref{m_voigt}) }{symbol:m_v}
\addnotation n: {porosity}{symbol:n}
\addnotation p: {pore pressure, (Eqn. \ref{zeta_press})}{symbol:p}
\addnotation \bold q: {specific discharge, (Eqn. \ref{discharge}) }{symbol:q}
\addnotation u_i: {i-th component of the displacements vector}{symbol:u_i}
\addnotation \bold v_f: {velocity of the fluid phase}{symbol:v_f}
\addnotation \bold v_s: {velocity of the solid phase}{symbol:v_s}

\addnotation : {}{symbol:}
\addnotation B: {Skempton's pore pressure coefficient, (Eqns. \ref{Bishop_B}, \ref{B_2})}{symbol:B}
\addnotation B': {pore pressure coefficient under uniaxial strain (or loading efficiency), (Eqns.  \ref{B_uniaxial}, \ref{B_uniaxial_3}, \ref{B_uniaxial_undrained},\ref{B_uniaxial2}) }{symbol:B'}
\addnotation C: {compressibility of the solid skeleton (drained compressibility of the porous medium) }{symbol:C}
\addnotation C_s: {compressibility of the solid phase (considered equal to the compressibility of solid grains), (Eqn.  \ref{compress_solidPhase})}{symbol:Cs}
\addnotation C_{\phi}: {unjacketed pore compressibility, (Eqn.  \ref{compress_pore})}{symbol:Cs}
\addnotation D: {elasticity stiffness matrix, Eqn. (\ref{D_matrix_3d})}{symbol:D}
\addnotation E: {Young's modulus}{symbol:E}
\addnotation E_u: {undrained Young's modulus, (Eqn. \ref{undrained_E})}{symbol:Eu}
\addnotation G: {shear modulus, (Eqn. \ref{shear_modulus})}{symbol:G}
\addnotation K: {bulk modulus, (Eqn. \ref{shear_modulus}) }{symbol:K}
\addnotation K_u: {undrained bulk modulus, (Eqns. \ref{k_undrained}, \ref{k_undrained_gassman}, \ref{undrained_shear_modulus}) }{symbol:Ku}
\addnotation K_f: {bulk modulus of the fluid phase }{symbol:Kf}
\addnotation K_s: {bulk modulus of the solid phase }{symbol:Ks}
\addnotation M: {Biot's modulus, (Eqn.  \ref{biot_modulus}, \ref{biot_M_ku})}{symbol:M}
\addnotation S: {uniaxial storage coefficient, (Eqns.  \ref{uniaxial_storage}, \ref{uni_storage_def},
                  \ref{coefficient_of_consolidation_3d}, \ref{uniaxial_storage_2})}{symbol:S}
\addnotation S_{\epsilon}: {storage coefficient, (Eqns. \ref{storativity_Cvoid}, \ref{storativity},  \ref{storativity_3}) }{symbol:Se}
\addnotation S_r: {degree of saturation}{symbol:Sr}

\addnotation : {}{symbol:}
\addnotation \nabla_s: {symmetric gradient operator, Eqn. (\ref{symm_grad_3d}) }{symbol:nabla_symmetric}

\end{tabbing}
 \clearpage

\section{Introduction}

A description of the mechanical behaviour of fluid saturated porous media under the assumption of
small perturbations is presented. The treatment falls within the framework of Biot's theory 
of consolidation, thus it is phenomenological.
The instigator for the development of poroelasticity was the solution to the problem of soil consolidation
\footnote{ Excellent reviews of the initial developments in the theory of porous media have been written by \cite{deBoer1996, deBoer2000}}.
The first treatment of this problem was a phenomenological approach by \cite{terzaghi_1925, terzaghibook}, who
considered soil to be laterally confined, thus undergoing uniaxial deformations. In Terzaghi's approach both 
solid and fluid constituents of the porous medium are considered incompressible. 
A general three-dimensional theory of elastic deformation of fluid infiltrated porous media
was proposed by \cite{biot1941}, in which the limitation of incompressible 
constituents was removed. Furthermore,  the increment of fluid content per unit volume was 
introduced as a variable work conjugate to the pore pressure. 
In \cite{biot_anisotropy}, the theory was extended to the general anisotropic elastic case.
The equations for the dynamic response of porous media were derived in \cite{BiotDyn}, while
extensions to nonlinear elasticity were presented in \cite{biot_nonlin}.
A formulation of Biot's linear
theory suitable for problems of soil mechanics was proposed by \cite{verruijt1969}, while 
\cite{rice1976} reformulated the equations of consolidation in terms of undrained coefficients.
Thus, the distinction between drained and undrained description of the equations of consolidation was 
introduced. 
Extensions to the use of nonlinear constitutive law
were proposed among others by \cite{Zienkiewicz80} and \cite{prevost80,Prevost82}.
A general treatment of Biot's theory in the range of nonlinear material behaviour and large deformations can be found in
\cite{coussy1991poromechanics, coussy2004poromechanics}, where the theory is reformulated using a thermodynamics approach. \\

Fluid infiltrated porous media consist of solid skeleton and fluid material that occupies the 
porous space. The mechanical behaviour of such media accounts for the coupling of skeletal 
deformations and fluid behaviour. The material considered here is isotropic, 
 and undergoes quasi-static deformations under isothermal conditions. 
 Porosity refers only to the connected porous space - however,
 the existence of isolated voids or cracks within the solid skeleton is not excluded.
 Solid phase is compressible and is not necessarily composed of a single constituent
 \footnote{The effect of voids and cracks occluded within the solid skeleton, and the existence of
 multiple solid constituents are included through the consideration of the unjacketed pore compressibility
 in the storage and pore pressure coefficients.}.
Pore fluid is compressible and consists of a single phase - for example, it can be water containing 
isolated air bubbles. The range of applicability of this theory is typically for a degree of liquid 
saturation higher than $90\%$. For lower degrees of saturation in the range of $0.75-0.8$,
researchers have used Biot's theory still assuming that the air and liquid pressures are equal 
(see, for example, \cite{okusa85}). 

A fundamental principle used
in the description of the mechanical behaviour of porous media is the principle of effective stress. 
This principle describes the decomposition of internal stresses applied to a porous medium. Accordingly, 
part of the stresses applied are transmitted to the pore fluid and the rest are transmitted to the 
solid skeleton. The former component causes changes in pore pressure and subsequent fluid flow. The 
latter component is the effective part of stresses that causes deformations on the solid skeleton. 
Consequently, the equations of equilibrium are the same as in classical elasticity, only expressed in 
terms of effective stresses. As it is reasonable to expect, conservation of mass is also required for 
the mechanical description to account for fluid flow. Fluid flow is considered to be viscous and is 
governed by Darcy's law. The fluid phase manifests itself in the equations through pore pressure, $p$, 
or the increment of fluid content per unit volume, $\zeta$.
The use of the increment of fluid content in the equations of equilibrium is associated with undrained 
poroelastic constants and leads to the undrained description of the equations of poroelasticity.


The present text has been written having in mind readers aiming at a first introduction to the theory of 
poroelasticity - particularly engineers with a background in soil mechanics who would like to access 
the general literature in the field of poroelasticity. It aims at bridging the gap between the
formulation of poroelasticity as used in the field of soil mechanics, with the generality of the 
formulation of Rice and Cleary, favoured in problems of rock mechanics. The exposition is phenomenological. 
The point of departure are the basic equations of elasticity (i.e. constitutive law, equations of equilibrium 
in terms of stresses, and the definition of strain), together with the principle of effective stress, from 
which the poroelastic equations of equilibrium are derived. Next, the equation of fluid mass conservation 
is introduced as derived by Verruijt. The expression derived herein is slightly more general, as it also 
includes the effect of the unjacketed pore compressibility. So far pore pressure is used as an independent 
variable in the description. The only new coefficients that are necessarily introduced in the equations 
are Biot's coefficient that appears in the principle of effective stress, the storage coefficient that 
appears in the equation of mass conservation, and definitions of material compressibility. Next, the 
increment of fluid content is defined and its constitutive law is derived, following naturally from the 
equation of fluid mass conservation. Pore pressure, storativity, and undrained poroelastic coefficients 
are also introduced and useful relations are proven. Equilibrium and mass conservation are subsequently 
expressed in terms of the increment of fluid content and undrained coefficients, 
leading to Rice and Cleary's formalism. Lastly, a weak ($\mathbf u, p$) formulation of poroelastic problems 
is presented.

\newpage

\section{Basic equations of isotropic elasticity}

Before proceeding to the exposition of the theory  of consolidation, it is useful to first review  the basic
equations of elasticity - both for  compressible and incompressible elasticity.
The equations and derivations related to this section can be found, for example, 
in \cite{westergaard1952theory}.
The equations of the classical theory of elasticity can be fully defined using two material constants. 
An appropriate pair of elasticity constants can be  Young's modulus, $E$, and  Poisson's ratio, $\nu$. 
Other fundamental constants that can be used are the bulk modulus, $K$, the shear modulus, $G$, and 
Lam\'{e}'s constant,  $\lambda$, which are linked to  Poisson's ratio and Young's modulus by the relations 
%
%
 \begin{equation}
   K = \dfrac{E}{3(1-2\nu)}; \hspace{0.5cm} G = \dfrac{E}{2(1+\nu)}; \hspace{0.5cm} \lambda = \dfrac{E\nu}{(1+\nu)(1-2\nu)}
\label{shear_modulus}
\end{equation}
Further useful relations among elasticity constants are the following:
\begin{equation}
   \lambda = K - \dfrac{2}{3}G; \hspace{0.5cm} \lambda = G\dfrac{2\nu}{1-2\nu}; \hspace{0.5cm} K = G\dfrac{2(1+\nu)}{3(1-2\nu)}
\label{shear_modulus_2}
\end{equation}
Using index notation, the components of the strain tensor are defined as
\begin{equation}
 \epsilon_{ij} = \dfrac{1}{2}\left( \dfrac{\partial u_i}{\partial x_j} + \dfrac{\partial u_j}{\partial x_i}  \right)
 \label{strain}
\end{equation}
where $u_i$ is the $i^{th}$ component of displacement. The constitutive law  can be written in a strain-stress relation as
\begin{equation}
 \epsilon_{ij} = \dfrac{1+\nu}{E}\sigma_{ij} - \dfrac{3\nu}{E}\sigma \delta_{ij} 
 \label{stress_strain_const}
\end{equation}
where $\sigma$ is the mean stress $\sigma = \sigma_{kk}/3$ and $\delta_{ij}$ is  Kronecker's delta. 
Inverting (\ref{stress_strain_const}) the stress-strain relations are obtained as
\begin{equation}
 \sigma_{ij} = (K - \dfrac{2}{3}G) \epsilon\delta_{ij} + 2G \epsilon_{ij} 
 \label{strain_stress_const}
\end{equation}
with $\epsilon$ being the volumetric strain, which is the first invariant of the strain tensor defined as
$\epsilon = \nabla\cdot\mathbf{u}$.
From the above equation  the relation connecting  mean stress to  volumetric strain can be retrieved
\begin{equation}
   \sigma = K\epsilon
 \label{stress_strain_bulk}
\end{equation}
Furthermore, the equation of equilibrium in terms of stress reads
\begin{equation}
 \dfrac{\partial \sigma_{ij}}{\partial x_j} + b_i = 0
 \label{eq_stress}
\end{equation}
where $b_i$ is the $i^{th}$ component of the applied body forces.

\subsection{Compressible elasticity}

The equations of equilibrium (\ref{eq_stress}) can be expressed in terms of the displacements, making use of the 
constitutive law
(\ref{strain_stress_const}) and the kinematics equations (\ref{strain}). Substituting (\ref{strain_stress_const}) into 
the equilibrium equation (\ref{eq_stress}) results in

\begin{equation}
   2G\dfrac{\partial \epsilon_{ij} }{\partial x_j} +\lambda \dfrac{\partial\epsilon}{\partial x_i} + b_i = 0
 \label{eq_comp1}
\end{equation}
Furthermore,  $\epsilon_{ij}$ is given in (\ref{strain}), which, substituted into equations (\ref{eq_comp1}) yields
\begin{equation}
   G  \dfrac{\partial^2 u_i}{\partial x_j \partial x_j} +    
  G\dfrac{\partial}{\partial x_i}\left(\dfrac{\partial u_j} {\partial x_j}\right)  
   +\lambda \dfrac{\partial\epsilon}{\partial x_i} + b_i = 0
 \label{eq_comp2}
\end{equation}
Considering that the following relations hold
\begin{equation}
 \nabla^2(\cdot) = \dfrac{\partial^2(\cdot)}{\partial x_j \partial x_j} ; 
   \hspace{0.5cm} \epsilon = \dfrac{\partial u_j} {\partial x_j} 
  \label{nabla_epsilon_index}
\end{equation}
equations (\ref{eq_comp2}) are written as
\begin{equation}
   G  \nabla^2 u_i +    (G+\lambda)\dfrac{\partial \epsilon}{\partial x_i}      + b_i = 0
 \label{Navier_compressible}
\end{equation}

\subsection{Incompressible elasticity}
\label{sec:incompressible_elasticity}

In the limit of incompressible material behaviour, where $\nu = 1/2$, the equilibrium equations 
(\ref{Navier_compressible}) do not hold, since Lam\'{e}'s constant, $\lambda$, becomes infinite. 
To circumvent this barrier, the equations of equilibrium can  be formulated in terms of displacements 
and  mean stress, and thus remain valid in the incompressibility limit. 
The first step in retrieving a mean stress formulation is adopting the constitutive  relations 
\begin{equation}
 \sigma_{ij} = 2G\epsilon_{ij} +  \dfrac{3\nu}{1+\nu}\sigma \delta_{ij}, \hspace{0.2cm} and \hspace{0.5cm} \epsilon = \sigma/K
 \label{stress_strain_const_incompress1}
\end{equation}
in which the mean stress, $\sigma$, is viewed as an independent variable.
Relation (\ref{stress_strain_const_incompress1}) can be obtained by substituting 
(\ref{stress_strain_bulk}) into (\ref{strain_stress_const}). The difference in this case though, is that 
equations (\ref{stress_strain_const_incompress1}) are valid even for 
$\nu = 1/2$, which results to $K\rightarrow\infty$.
Substituting (\ref{stress_strain_const_incompress1}) into the equilibrium
 equations  (\ref{eq_stress}), yields
\begin{equation}
   2G\dfrac{\partial \epsilon_{ij} }{\partial x_j} + \dfrac{3\nu}{1+\nu}\dfrac{\partial\sigma}{\partial x_i} + b_i = 0
 \label{eq_incomp1}
\end{equation}
or
\begin{equation}
   G  \dfrac{\partial^2 u_i}{\partial x_j \partial x_j} +    
  G\dfrac{\partial}{\partial x_i}\left(\dfrac{\partial u_j} {\partial x_j}\right)  
   + \dfrac{3\nu}{1+\nu}\dfrac{\partial\sigma}{\partial x_i} + b_i = 0
 \label{eq_incomp2}
\end{equation}
which due to (\ref{nabla_epsilon_index}) become
\begin{equation}
   G  \nabla^2u_i +    
  G\dfrac{\partial}{\partial x_i}\epsilon
   + \dfrac{3\nu}{1+\nu}\dfrac{\partial\sigma}{\partial x_i} + b_i = 0, \hspace{0.5cm} \epsilon = \sigma/K
 \label{eq_incomp3}
\end{equation}
In the limit of material incompressibility, where $\nu=1/2$, the bulk modulus tends to infinity ($K\rightarrow\infty$). 
Since stresses are finite, then relation $\epsilon = 0$ should hold. Therefore, for incompressible material behaviour, 
and using vector notation, the equations of equilibrium  become
\begin{equation}
  G  \nabla^2 \mathbf{u} + \nabla\sigma + \mathbf{b} = 0,\hspace{0.2cm} and \hspace{0.5cm}  \nabla\cdot \mathbf u = 0
  \label{Navier_incompressible}
\end{equation}
$\nabla\cdot \mathbf u = 0$ expresses the incompressibility condition.

\newpage

\section{Principle of effective stress}
\label{sec:effective_stress}

Stresses applied to a saturated porous medium
are partly distributed to the solid skeleton and partly to the pore fluid. 
The former stresses are responsible for skeletal deformations, this is why they are called effective.
Considering that stresses are positive when they are tensile and pressure is positive when it is compressive, 
the principle of effective stress is written in index notation as
\begin{equation}
  \sigma_{ij} = \sigma'_{ij} - \alpha p \delta_{ij}
\label{eff_stress}
\end{equation}
In equation (\ref{eff_stress}), $\sigma_{ij}$ and $\sigma'_{ij}$ are the components of the total and effective stress 
and $p$ is the pore pressure. The symbol $\delta_{ij}$ is  Kronecker's delta, defined as  
$\delta_{ij}=1$ for $i=j$, and $\delta_{ij}=0$ for $i\neq j$. 
The parameter
$\alpha$ is known as Biot's coefficient. In the work of \cite{terzaghibook}, this parameter was considered to have the
value of one - an assumption generally valid for soil.

\subsection*{Biot's coefficient $\alpha$}
 \cite{biot1941} expressed coefficient $\alpha$ of equation (\ref{eff_stress})  as
\begin{equation}
 \alpha = \dfrac{K}{H}
\label{alpha}
\end{equation}
where $K$ is the drained bulk modulus of the porous material, and $1/H$ is the poroelastic expansion coefficient, that was 
 introduced by Biot. It describes the change 
of the bulk volume due to a pore pressure change while the stress is constant.
\cite{BiotWillis} recast the above equation in terms of two coefficients of 
unjacketed and jacketed compressibility.
 The unjacketed coefficient 
 of compressibility is the compressibility of the solid phase $C_s$
 \footnote{The compressibility of the solid phase is often considered identical to the compressibility of the grains. This, however, is 
 true if the skeleton is composed of one mineral (\cite{wang2000}).}
 and the 
jacketed compressibility coefficient is the drained compressibility of the porous material $C$
\footnote{The unjacketed compressibility is measured in an undrained test (see for example \cite{wang2000}, Section 3.1.2,
or \cite{BiotWillis}).
To perform the test a sample is immersed in  fluid under pressure, with the fluid penetrating the pores of the material. 
Any change in the applied confining pressure produces an  equal change to the pore pressure. Denote the confining pressure 
with $\Delta P_c$ and pore pressure with $\Delta p$, then for this experiment the condition $\Delta P_c = \Delta p$ holds. 
The compressibility of the solid phase is calculated  as 
 
 \begin{equation*}
   C_s = - \dfrac{1}{V} \dfrac{\Delta V}{\Delta p}|_{\Delta P_c = \Delta p}
 \end{equation*}

The jacketed compressibility is 
measured in a drained test, where the sample is covered by a surface membrane and the inside of the jacket is connected to
the atmosphere using a tube. The tube connection to the atmosphere makes the pore pressure remain constant under the load
applied to the specimen.}.
In familiar notation of soil mechanics the expression they provided is

\begin{equation}
 \alpha = 1 - \frac{C_s}{C}
\label{alpha2}
\end{equation}
The above relation has also been derived independently from Bishop and Skempton (\cite{SkemptonBook}).
A derivation can also be found in \cite{bishop1973}.\\
For soft soils, the value of $\alpha$ is considered to be one and the principle of effective stress reduces to 
\begin{equation}
  \sigma_{ij} = \sigma'_{ij} -  p \delta_{ij}
\label{eff_stress_terzaghi}
\end{equation} 
This is the expression used by Terzaghi in his original work and is valid for most soils. Soils are usually soft and their skeleton is
highly compressible, while their particles (solid phase) have small compressibility, which justifies the use of the coefficient 
$\alpha$ as equal to 1. In contrast, this is not always the case for rocks. Table \ref{tab:Biot_Willis_coefficient} 
presents typical values of Biot's coefficient and is compiled with data obtained from  \cite{mitchell2005fundamentals}.

\begin{table}[h!]
\caption{ Biot's coefficient for Soil and Rock materials.}
\centering
 \begin{tabular}{l*{2}{c}r}
 \hline
material                        & $C_s/C$ & $\alpha$\\
\hline
Dense sand                      & 0.0015 & 0.9985  \\
Loose sand                      & 0.0003 & 0.9997  \\
London clay (over cons.)        & 0.00025 & 0.99975   \\
Gasport clay (normally cons.)   & 0.00003 & 0.99997   \\
Quartzitic sandstone            & 0.46    & 0.54  \\
Quincy granite (30 m deep)      & 0.25    & 0.75   \\
Vermont marble                  & 0.08    & 0.92   \\
\hline     \\
Data obtained from \cite{mitchell2005fundamentals}.
\end{tabular}
\label{tab:Biot_Willis_coefficient}
\end{table}

\section{Drained description of poroelastic equations}
\label{sec:drained_behaviour}

Two limiting regimes of deformation  define the consolidation of porous media, drained deformations and undrained deformations.
Drained deformations take place under constant fluid pore pressure, while during undrained deformations no fluid flux is 
permitted on the boundaries of the control volume, which means that undrained deformations take place under constant fluid
mass content. The poroelastic behaviour of concern to this work falls  between these two limiting behaviours. Undrained
behaviour is examined as well, though, since it leads to the definition of undrained constants (see Section \ref{undrained_parameters}).

\subsection{Constitutive relations}
\label{sec:constitutive_drained}

The stress-strain relationships  for fluid saturated porous media  are identical to the ones 
of nonporous media, provided that they are expressed in terms of the effective stress as dictated by the principle of effective stress. 
In a compliance formulation this is  presented as
\begin{equation}
\epsilon_{ij}  =  \dfrac{1+\nu}{E} \sigma'_{ij} - \dfrac{\nu}{E}\sigma'_{kk}\delta_{ij}  
\label{3d_constitutive_law}
\end{equation}
Equations (\ref{3d_constitutive_law}) differ from (\ref{stress_strain_const}) in that stresses $\sigma'_{ij}$ are 
the effective stresses applied on the soil skeleton. 
From equations (\ref{3d_constitutive_law}) the equivalent stiffness formulation can be derived as
\begin{equation}
\sigma'_{ij}  =  \left(K - \dfrac{2G}{3}\right)\epsilon\delta_{ij} +   2G\epsilon_{ij}
\label{stress_strain_eff_3d}
\end{equation}
where $\epsilon$ is the volumetric strain, $\epsilon = \epsilon_{kk}$ (indices appearing twice indicate summation under
Einstein's convention).
Making use of the principle of effective stress (equation \ref{eff_stress}), the stress-strain relations 
(\ref{3d_constitutive_law}) and (\ref{stress_strain_eff_3d}) are expressed in terms of the total stresses and pore pressure as
\begin{equation}
\epsilon_{ij}  =  \dfrac{1}{2G} \left(\sigma_{ij} - \dfrac{\nu}{1 + \nu}\sigma_{kk}\delta_{ij} \right) 
              + \dfrac{\alpha}{3K} p\delta_{ij} 
\label{strain_stress_total_3d}
\end{equation}
and
\begin{equation}
\sigma_{ij}  =  \left(K - \dfrac{2G}{3}\right)\epsilon\delta_{ij} +   2G\epsilon_{ij} - \alpha p \delta_{ij}
\label{stress_strain_total_3d}
\end{equation}
From equation (\ref{stress_strain_eff_3d}), the isotropic (or mean) effective stress can be expressed with respect 
to the bulk modulus of the porous material  and the volumetric 
strain as 
\begin{equation}
 \dfrac{\sigma'_{kk}}{3} = K\epsilon
\label{mean_eff_stress_1}
\end{equation}
which, in terms of the compressibility and  mean effective stress, is expressed as
\begin{equation}
 \sigma' = \dfrac{\epsilon}{C}
 \label{mean_eff_stress_1_2}
\end{equation}
Making use of the principle of effective stress (\ref{eff_stress}), equation (\ref{mean_eff_stress_1})
can be written in terms of  mean stress as
\begin{equation}
 \sigma + \alpha p = K\epsilon
\label{mean_eff_stress_2}
\end{equation}
Last, stress-strain relationship with regard to mean stress can be derived by substituting the latter equation into equation
(\ref{stress_strain_total_3d}) and making use of the relations (\ref{shear_modulus_2})  thus reads
\begin{equation}
\sigma_{ij}  =    2G\epsilon_{ij} + \dfrac{3\nu}{1+\nu}\sigma\delta_{ij}  - 2G\dfrac{\alpha p}{3K} \delta_{ij}
\label{stress_strain_mean_stress}
\end{equation}
Therefore, the constitutive relation (\ref{stress_strain_total_3d}) can be substituted by the two relations
(\ref{stress_strain_mean_stress}) and (\ref{mean_eff_stress_2}).

\subsection*{Coefficient of one dimensional compressibility}
Assume that a column of fluid infiltrated porous material is colinear with the z-axis, and cannot deform lateraly.
Denoting the effective  stresses on the z-direction with   $\sigma'_{zz}$, and corresponding soil strains with 
$\epsilon_{zz}$, the one-dimensional constitutive law for the soil is written as 
\begin{equation}
\epsilon(z)  =  m_v \sigma'(z)  
\label{constitutive_law}
\end{equation}
The term $m_v$ in equation (\ref{constitutive_law}) represents the drained (i.e. $p = 0$) vertical compressibility of the laterally 
confined soil. The value of $m_v$ can be calculated using equations (\ref{3d_constitutive_law}). 
Considering the boundary conditions of the one dimensional consolidation problem
which allow only for vertical frictionless displacements, shear strains and horizontal strains 
are set to zero. Vanishing shear strains lead 
to shear stresses being zero. Furthermore, substituting $\epsilon_{x} = 0$
and $\epsilon_{y} = 0$ into (\ref{3d_constitutive_law}) yields
\begin{equation}
\nu(\sigma'_x + \sigma'_y)  =  \dfrac{2\nu^2}{1-\nu} \sigma'_{z}  \\
\label{3d_stresses_relationship}
\end{equation}
Substituting equation (\ref{3d_stresses_relationship}) to equation (\ref{3d_constitutive_law}) and for 
$i = j = z$, results to
\begin{equation}
\epsilon_{zz}  =  \dfrac{(1+\nu)(1-2\nu)}{E(1-v)} \sigma'_{zz} \\
\label{3d_constitutive_law_2}
\end{equation}
which makes the value of $m_v$ equal to
\begin{equation}
\begin{array}{r l l}
           m_v &\hspace{-0.2cm} =&  \hspace{-0.2cm} \dfrac{(1+\nu)(1-2\nu)}{E(1-\nu)}   \\      
 \end{array} 
\label{volume_change}
\end{equation}
or
\begin{equation}
\begin{array}{r l l}
     m_v &\hspace{-0.2cm}=& \hspace{-0.2cm} \dfrac{1}{(\lambda + 2G)}    \\
  \end{array} 
\label{volume_change1}
\end{equation}
In equations (\ref{volume_change}) and (\ref{volume_change1}), $E$ and $\nu$ are the Young's 
modulus and the Poisson ratio of the porous medium after the excess water is squeezed out (drained), respectively,
$G$ is the shear modulus, and $\lambda$ is a Lam\'{e} constant (drained).

\subsection{Equations of equilibrium}
\label{sec:equilibrium_eqns}

In this section, the  equations of equilibrium of fluid saturated porous medium in terms of 
 skeleton displacements and the pore pressure  are derived. 
The equations are presented in  matrix notation, but where appropriate they are stated using index notation as well. 
The presentation of the equations of equilibrium in matrix notation  proves convenient in Section \ref{sec:weak_form}
were the weak form of the equations of consolidation is derived. 

Stresses and strains are represented using Voigt notation for symmetric tensors. In three dimensions 
they read 
\begin{equation}
\begin{aligned}
\mathbf{\sigma} &= [ \sigma_{xx},\sigma_{yy}, \sigma_{zz}, \sigma_{xy}, \sigma_{yz}, \sigma_{zx} ]^T
\\
\mathbf{\epsilon} &= [ \epsilon_{xx}, \epsilon_{yy}, \epsilon_{zz}, \gamma_{xy}, \gamma_{yz}, \gamma_{zx} ]^T
\end{aligned}
\label{sigma_voigt}
\end{equation}
with $\gamma_{ij} = 2\epsilon_{ij}$ for $i\neq j$. The principle of effective stress is now written as 
\begin{equation}
 \mathbf{\sigma} = \mathbf{\sigma'} - \alpha\mathbf m p
\label{effective_stress_matrix}
\end{equation}
(\cite{zienkiewicz2005finite,zienkiewiczBook}), where $\mathbf m$ is the vector form of Kronecker's delta, $\delta_{ij}$. 
In three dimensions $\mathbf m$ reads
\begin{equation}
\mathbf{m} = [ 1,1, 1, 0, 0, 0 ]^T
\label{m_voigt}
\end{equation}
Furthermore, the linear constitutive law is given by 
\begin{equation}
 \mathbf\sigma' = D\mathbf\epsilon
\label{constitutive}
\end{equation}
The matrix $D$ in elasticity theory is often called the stiffness matrix and in three dimensions it is equal to
\begin{equation}
 D = \dfrac{E}{(1+\nu)(1-2\nu)} \begin{bmatrix} 1-\nu & \nu & \nu & 0 & 0 & 0 
 \\ \nu & 1-\nu & \nu & 0 & 0 & 0 \\ \nu & \nu & 1-\nu & 0 & 0 & 0 
 \\ 0 & 0 & 0 & \dfrac{1-2\nu}{2} &0 & 0 \\ 0 & 0 & 0 & 0 & \dfrac{1-2\nu}{2} &0  
\\ 0 & 0 & 0 & 0 & 0 & \dfrac{1-2\nu}{2} \end{bmatrix}
\label{D_matrix_3d}
\end{equation}
Lastly, strains and displacements are linked through the kinematics relations 
\begin{equation}
 \mathbf\epsilon = \nabla_s\mathbf u
 \label{kinematix_voigt}
\end{equation}
with $\mathbf u = [u, v, w]^T$.
$\nabla_s$ appearing in the above equation denotes the symmetric gradient operator, which for  three dimensional problems 
is defined as
\begin{equation}
\nabla_s = \begin{bmatrix}  \partial/\partial x & 0 & 0\\  0 & \partial/\partial y & 0 \\  0 & 0 & \partial/\partial z
\\ \partial/\partial y & \partial/\partial x & 0 \\ 0 & \partial/\partial z & \partial/\partial y 
\\   \partial/\partial z & 0 & \partial/\partial x  \end{bmatrix}
\label{symm_grad_3d}
\end{equation}
Using (\ref{kinematix_voigt}), equation (\ref{constitutive}) gives
\begin{equation}
 \mathbf\sigma' = D\nabla_s\mathbf u
\label{constitutive1}
\end{equation}
In the presence of body forces, the equilibrium equations are given by  
\begin{equation}
 \nabla^T_s\mathbf\sigma + \mathbf b = 0
\label{equilibrium_matrix_form1}
\end{equation}
 which in index notation is written as
\begin{equation}
\dfrac{\partial\sigma_{ij}}{\partial x_j} + b_i = 0
\label{equilibrium_index}
\end{equation}
Assuming that only  gravitational body forces are applied, $\mathbf b$ is defined as $\mathbf b = [0,0,\rho g]^T$.
In the definition of $\mathbf b$,
 $g$ is the acceleration of gravity, and $\rho$ is the average density of the porous seabed given by the 
equation $\rho = \rho_f n +\rho_s (1-n)$. In the latter expression, $\rho_f$ is the fluid density, $\rho_s$ 
is the soil density, and $n$ is the soil porosity.

Substituting the matrix form of the effective stress principle (\ref{effective_stress_matrix}) in
 equation (\ref{equilibrium_matrix_form1}) results to the  equilibrium equations in terms of stresses
\begin{equation}
 \nabla^T_s\sigma' - \nabla^T_s(\alpha \mathbf m \rho) + \mathbf b = 0
\label{equilibrium_stress_form_vector}
\end{equation}
or in index notation
\begin{equation}
 \dfrac{\partial\sigma'_{ij}}{\partial x_j} - \alpha \delta_{ij}\dfrac{\partial\rho}{\partial x_i} +  b_i = 0
\label{equilibrium_stress_form}
\end{equation}
The substitution of equation (\ref{constitutive1}) in the last equation, leads to the equilibrium equations in term 
of displacements
\begin{equation}
 \nabla^T_sD\nabla_s\mathbf u = \nabla^T_s(\alpha \mathbf m \rho) - \mathbf b
\label{equilibrium_matrix_form}
\end{equation}
Using index notation, the equations of equilibrium (\ref{equilibrium_matrix_form}) take the form 
\begin{equation}
     \dfrac{E}{2(1+\nu)}\left( \nabla^2 u_i + \dfrac{1}{1-2\nu} \dfrac{\partial\epsilon}{\partial x_i}
                 \right) - \alpha\dfrac{\partial p}{\partial x_i} = -b_i 
 \label{equilibrium_final_3d}
\end{equation}
with $\epsilon = \frac{\partial u}{\partial x} + \frac{\partial v}{\partial y} + \frac{\partial w}{\partial z}$ 
being the volumetric strain in three dimensions. 
Introducing the shear modulus and Lam\'{e}'s constant $\lambda$, as defined in equations ($\ref{shear_modulus}_1$) and 
($\ref{shear_modulus_2}_2$), 
 the equations of equilibrium (\ref{equilibrium_final_3d}) are written as 

\begin{equation}
  G \nabla^2 u_i + \left(\lambda + G\right) \dfrac{\partial\epsilon}{\partial x_i}
                 = \alpha\dfrac{\partial p}{\partial x_i}  -b_i
\label{navier_index_1}
\end{equation}
The system of equations (\ref{navier_index_1})  contains one variable in excess.
An additional equation is required to complement the boundary value problem, and it is obtained from  the 
conservation of fluid mass, which is examined in section \ref{mass_conservation}.

\subsection{Equations of compatibility}

A problem of isothermal  poroelasticity  is fully defined 
using the equilibrium equations  expressed in terms of displacements and  an additional equation  expressing 
the conservation of mass. The equations of equilibrium can also be formulated  with respect to stresses, as 
in equation (\ref{equilibrium_stress_form}). In this case the set of unknown variables 
\footnote{Which are the stresses and the pore pressure.}
is larger than the available equations of equilibrium and mass continuity. One more equation is required 
for plane strain problems, and three more equations are required for three dimensional elasticity problems, due 
tothe fact that the displacement field should satisfy certain continuity requirements. The additional 
equations to solve the poroelasticity problem are obtained by the kinematic compatibility equations, which when 
expressed in terms of strain are called the Saint-Vainant compatibility equations. The compatibility equations 
can be expressed as well in terms of stresses, in which case they are called Beltrami-Michell equations. 
In the 
following, departing from the strain compatibility conditions, the Beltrami-Michell equations are derived  
for the plane strain and three dimensional poroelasticity.

\subsubsection{Kinematic compatibility in plane strain conditions}
\label{compatibility_2d}
For the case of plane strain there is one kinematic compatibility equation, which is first  derived in terms of strain. 
The strain-displacement relations for plane strain are 
\begin{equation}
    \epsilon_{xx}  =  \dfrac{\partial u}{\partial x}; \hspace{0.5cm}    
     \epsilon_{zz}  =  \dfrac{\partial w}{\partial z};  \hspace{0.5cm}
     \gamma_{xz}  =  \dfrac{\partial u}{\partial z} + \dfrac{\partial w}{\partial x}
\label{plane_strain_eq}
\end{equation}
Eliminating the displacements from equations (\ref{plane_strain_eq}) leads to the strain compatibility equation 
\begin{equation}
\dfrac{\partial^2\epsilon_{xx}}{\partial z^2} + \dfrac{\partial^2\epsilon_{zz}}{\partial x^2} = 
      \dfrac{\partial^2\gamma_{xz}}{\partial x \partial z}
\label{strain_compatibility_eq}
\end{equation}
The Beltrami-Michell equation of  compatibility in terms of stress is derived next, which is achieved 
transforming equation (\ref{strain_compatibility_eq}) to an equation of the stresses $\sigma'_{xx}$,  
$\sigma'_{zz}$, and the pressure $p$.  
For plane strain conditions, $\epsilon_{yy} = 0$, and using the constitutive relations  (\ref{3d_constitutive_law}),
 leads to the condition
\begin{equation}
 \sigma'_{yy} = \nu\left(\sigma'_{xx} + \sigma'_{zz}   \right) 
 \label{plane_strain_sigmay}
\end{equation}
Furthermore the shear stresses on the $x-y$ and $z-y$ directions are zero, because the equivalent strains are zero also.
Substituting equation (\ref{plane_strain_sigmay}) to $(\ref{3d_constitutive_law}_1$) and  
$(\ref{3d_constitutive_law}_2$), the expressions for $\epsilon_{xx}$ and $\epsilon_{zz}$ become
\begin{equation}
\begin{array}{r l l}
    \epsilon_{xx}  =&  \dfrac{1}{E} \left((1-\nu^2)\sigma'_{xx} - (\nu+\nu^2 )\sigma'_{zz}  \right) \\
 \\     
      \epsilon_{zz}  =&  \dfrac{1}{E} \left((1-\nu^2)\sigma'_{zz} - (\nu+\nu^2 )\sigma'_{xx}  \right)
  \end{array} 
\label{2d_constitutive_law_components_1_2}
\end{equation}
Next the shear stress $\sigma_{xz}$ is expressed with respect to $\sigma'_{xx}$,  
$\sigma'_{zz}$, and  $p$, using the equilibrium equations (\ref{equilibrium_index}), which for plane strain conditions
 are expanded as
\begin{equation}
\begin{array}{r l l}
   \dfrac{\partial\sigma'_{xx}}{\partial x} + \dfrac{\partial\sigma_{xz}}{\partial x} 
              - \alpha\dfrac{\partial p}{\partial x} = & -b_x  \\
 \\     
   \dfrac{\partial\sigma_{xz}}{\partial z} + \dfrac{\partial\sigma'_{zz}}{\partial z} 
              - \alpha\dfrac{\partial p}{\partial z} = & -b_z  \\
 \end{array} 
\label{plane_strain_eq_stress}
\end{equation}
Derivating the first of the equations ($\ref{plane_strain_eq_stress}$) with respect to $x$, the second with respect to $z$, and 
adding them, results to 
\begin{equation}
 2\dfrac{\partial^2\sigma_{xz}}{\partial x\partial z} = -\left( \dfrac{\partial^2\sigma'_{xx}}{\partial x^2} 
              + \dfrac{\partial^2\sigma'_{zz}}{\partial z^2} - \alpha\nabla^2p -\nabla\cdot\mathbf b\right)
 \label{plane_str_sh1}
\end{equation}
where $\mathbf b = (b_x, b_z)^T$ and $\nabla\cdot\mathbf b = \partial b_x/\partial x + \partial b_z/\partial z$. 
Shear stress-strain relation is
\begin{equation*}
 \gamma_{xz} = \dfrac{2(1+\nu)}{E}\sigma_{xz}
\end{equation*}
which under equation (\ref{plane_str_sh1}) becomes
\begin{equation}
 \dfrac{\partial^2\gamma_{xz}}{\partial x \partial z} = -\dfrac{1+\nu}{E}\left( \dfrac{\partial^2\sigma'_{xx}}{\partial x^2} 
              + \dfrac{\partial^2\sigma'_{zz}}{\partial z^2} -\alpha \nabla^2 p -\nabla\cdot\mathbf b\right)
 \label{plane_str_component3}
\end{equation}
The Beltrami-Michell equation for plane strain is now obtained substituting equations (\ref{2d_constitutive_law_components_1_2})
and (\ref{plane_str_component3}) to the strain compatibility equation (\ref{strain_compatibility_eq}), which in terms of the effective
streses reads  
\begin{equation}
 \nabla^2\left(\sigma'_{xx} + \sigma'_{zz} - \dfrac{\alpha p}{1-\nu} \right) = -\dfrac{1}{1-\nu}\nabla\cdot\mathbf b
 \label{plane_compatibility_1}
\end{equation}
In terms of total stresses the compatibility equation takes the form
\begin{equation}
 \nabla^2\left(\sigma_{xx} + \sigma_{zz} + 2\eta p \right) = -\dfrac{1}{1-\nu}\nabla\cdot\mathbf b
 \label{plane_compatibility}
\end{equation}
where $\eta$ is the poroelastic stress coefficient 
\begin{equation*}
\eta = \dfrac{\alpha(1-2\nu)}{2(1-\nu)}
 \label{poro_stress_coeff}
\end{equation*}
Compatibility equation (\ref{poro_stress_coeff}) associated with equations (\ref{plane_strain_eq_stress})
and the equation of mass continuity, can be used as a full set of equations to solve plane strain poroelastic 
problems. This formulation is suitable when only stress boundary conditions are applied. It should 
be noted at this point that for multiply connected domains, compatibility conditions are necessary but 
not sufficient condition to guarantee single-valued (continuous) displacements. For a discussion on this 
topic please refer to Section 10.4 of \cite{chou1992elasticity}.

\subsubsection{Kinematic compatibility in three-dimensional poroelasticity}
\label{compatibility_3d}

The equations of strain compatibility in three dimensional elasticity  are derived in a process similar 
to the derivation of equation (\ref{strain_compatibility_eq}) 
%
%
It is found 
that they form the system of six equations
\begin{equation}
 \dfrac{\partial^2\epsilon_{xx}}{\partial y^2} + \dfrac{\partial^2\epsilon_{yy}}{\partial x^2} = 
      \dfrac{\partial^2\gamma_{xy}}{\partial x \partial y} 
 \label{3d_stress_compatibility_1}
\end{equation}
\begin{equation}
 \dfrac{\partial^2\epsilon_{yy}}{\partial z^2} + \dfrac{\partial^2\epsilon_{zz}}{\partial y^2} = 
      \dfrac{\partial^2\gamma_{yz}}{\partial y \partial z} 
\end{equation}
\begin{equation}
 \dfrac{\partial^2\epsilon_{zz}}{\partial x^2} + \dfrac{\partial^2\epsilon_{xx}}{\partial z^2} = 
      \dfrac{\partial^2\gamma_{zx}}{\partial z \partial x}
\end{equation}
\begin{equation}
  2\dfrac{\partial^2\epsilon_{xx}}{\partial y \partial z} = \dfrac{\partial}{\partial x}
  \left( -\dfrac{\partial\gamma_{yz}}{\partial x} + \dfrac{\partial\gamma_{xz}}{\partial y} 
             + \dfrac{\partial\gamma_{xy}}{\partial z}    \right)
\end{equation}
\begin{equation}
  2\dfrac{\partial^2\epsilon_{yy}}{\partial z \partial x} = \dfrac{\partial}{\partial y}
  \left( \dfrac{\partial\gamma_{yz}}{\partial x} - \dfrac{\partial\gamma_{xz}}{\partial y} 
             + \dfrac{\partial\gamma_{xy}}{\partial z}    \right)
\end{equation}
\begin{equation}
  2\dfrac{\partial^2\epsilon_{zz}}{\partial x \partial y} = \dfrac{\partial}{\partial z}
  \left( \dfrac{\partial\gamma_{yz}}{\partial x} + \dfrac{\partial\gamma_{xz}}{\partial y} 
             - \dfrac{\partial\gamma_{xy}}{\partial z}    \right)
 \label{3d_stress_compatibility_6}
\end{equation}
The six compatibility equations (\ref{3d_stress_compatibility_1})-(\ref{3d_stress_compatibility_6}) are equivalent to 
three independent equations of fourth order \citep{chou1992elasticity}. 
Substituting the stress-strain equations (\ref{3d_constitutive_law}) and the equilibrium equations (\ref{equilibrium_index}) 
in the compatibility equations, results to the Beltrami-Michell compatibility equations 
(\cite{Detournay_Cheng1993}, p.31; \cite{wang2000}, p.77). 
Using index notation,
the resulting equations are expresses in terms of the effective stresses as
\begin{equation}
 \nabla^2\sigma'_{ij} + \dfrac{1}{1+\nu}\dfrac{\partial^2\sigma'_{kk}}{\partial x_i \partial x_j} 
 -\alpha\left[2\dfrac{\partial^2p}{\partial x_i \partial x_j} + \dfrac{\nu}{1-\nu}\delta_{ij}\nabla^2p \right] = 
            -\dfrac{\nu}{1-\nu}\delta_{ij}\nabla\cdot\mathbf b -\dfrac{\partial b_i}{\partial x_j} -\dfrac{\partial b_j}{\partial x_i}
 \label{beltrami_michell_1}
\end{equation}
where $\mathbf b = (b_x, b_y, b_z)^T$.
A second form which equation (\ref{beltrami_michell_1}) can take using total stresses, is 
\begin{equation}
 \nabla^2\sigma_{ij} + \dfrac{1}{1+\nu}\dfrac{\partial^2\sigma_{kk}}{\partial x_i \partial x_j} 
 +2\eta\left[\dfrac{1-\nu}{1+\nu}\dfrac{\partial^2p}{\partial x_i \partial x_j} + \delta_{ij}\nabla^2p \right] = 
    -\dfrac{\nu}{1-\nu}\delta_{ij}\nabla\cdot\mathbf b -\dfrac{\partial b_i}{\partial x_j} -\dfrac{\partial b_j}{\partial x_i}
 \label{beltrami_michell}
\end{equation}
where the coefficient $\eta$ is the same as in (\ref{poro_stress_coeff}). Last, contracting equations (\ref{beltrami_michell})
by setting $i=j$, results to the useful equation
\begin{equation}
 \nabla^2\left(\sigma_{kk} + 4\eta p \right) = -\dfrac{1+\nu}{1-\nu}\nabla\cdot\mathbf b
 \label{stress_compatibility_3d}
\end{equation}

\subsection{Darcy's law}
\label{sec_darcy}

In Biot's theory of consolidation  fluid flow is assumed to be governed by Darcy's law of fluid flow in
porous media. The general expression of the three dimensional Darcy's law is 
\begin{equation}
 \mathbf q = -  \dfrac{k}{\gamma_w}\nabla( p - \rho_f z g)
\label{darcy_0}
\end{equation}
(\cite{wang2000}), where $k$ is the specific permeability (or hydraulic conductivity), 
and $\gamma_w$ is the specific
weight of the fluid, defined as $\gamma_w = \rho_fg$. The vector $\mathbf q$ is specific discharge, that is, 
the relative velocity of the fluid component 
of a porous medium, $\mathbf v_f$, with respect to the velocity of the solid component, $\mathbf v_s$, multiplied by the 
porosity, $n$, and is represented as
\begin{equation}
 \mathbf{q} = n(\mathbf v_f - \mathbf v_s)
\label{discharge}
\end{equation}
%


\subsection{Fluid mass balance}
\label{mass_conservation}

Assuming that the consolidation process takes place under isothermal conditions, the equation of mass 
conservation for the infiltrating fluid is the only additional equation 
required to define the consolidation process. The equation of mass conservation for fluid-infiltrated 
poroelastic media is presented in this subsection. This derivation is an extension to that of
\cite{verruijt2008, verruijt_2013}.

Consider a constant mass quantity $m$ occupying in the current configuration volume $V$ with porosity $n$. 
The mean value of the velocity of fluid is $\mathbf{v_f}$, and that of solid is $\mathbf{v_s}$. 
We consider the mass conservation of the fluid 
filling the pores. Denoting the fluid density with $\rho_f$, the fluid mass filling the
pores of the saturated medium is $m_f = n\rho_fV$, and the relative fluid density is $n\rho_f$. 
In the absence of a source generating fluid mass, Eulerian fluid continuity equation reads
\begin{equation}
 \dfrac{\partial (n\rho_f)}{\partial t}   + \bigtriangledown \cdot (n\rho_f \mathbf v_f) = 0
\label{mass_balance}
\end{equation}
(\cite{coussy2004poromechanics}, \cite{Rudnicki_book}). 
Fluid compressibility, $\beta$, is related to the change in fluid pressure and the fractional change in fluid volume as 
$ \beta = - \dfrac{1}{V_f} \dfrac{\Delta V_f}{\Delta p}$, which leads to the constitutive relation 
\footnote{This relation can be shown as follows. Consider constant fluid mass, $m_f$. Mass continuity dictates that $\rho_{fo}V_{fo} = \rho_f V_f$, where $\rho_{fo}$ and $V_{fo}$ are
fluid density and volume at reference configuration and $\rho_{f}$ and $V_{f}$ are the equivalent quantities in the current configuration. Substituting
for $V_f = V_{fo} + \Delta V_f$ and $\rho_f = \rho_{fo} + \Delta\rho_f$ into the above equation of fluid mass continuity yields
$\dfrac{\Delta V_f}{V_{fo}} = -\dfrac{\Delta \rho_f}{\rho_f}$. In the range of small perturbations this can be written as

\begin{equation*}
 \dfrac{\partial V_f}{V_f} = -\dfrac{\partial \rho_f}{\rho_f}
\end{equation*}
}

\begin{equation}
 \dfrac{ \partial\rho_f}{\partial t}  = 
  \rho_f\beta \dfrac{\partial p}{\partial t} 
\label{fluid_compressibility_constitutive_1}
\end{equation}
Multiplying equation (\ref{fluid_compressibility_constitutive_1}) by $n$, and  substituting 
the resulting equation  to the fluid mass conservation equation (\ref{mass_balance}) leads to 
\begin{equation}
 \dfrac{\partial n}{\partial t} + n\beta\dfrac{\partial p}{\partial t} + \nabla \cdot(n\mathbf v_f) = 0
\label{fluid_mass_conservation_0}
\end{equation}
which can also be written as

\begin{equation}
 \dfrac{\partial n}{\partial t} + n\beta\dfrac{\partial p}{\partial t} + \nabla \cdot[n(\mathbf v_f -\mathbf v_s)] + \nabla \cdot(n\mathbf v_s) = 0
\label{fluid_mass_conservation}
\end{equation}
The term $n(\mathbf v_f -\mathbf v_s)$ in the above equation is identified with the specific discharge, $\mathbf q$, while the term 
$\nabla \cdot(n\mathbf v_s)$ is calculated using the conservation of mass of the solid skeleton.

Consider now the mass balance of the skeleton. Denoting the density with $\rho_s$, 
the solid mass is equal to $m_s = (1-n)\rho_sV$, and the relative density is equal to $(1-n)\rho_s$.
Same as with equation (\ref{mass_balance}), mass balance is written as
\begin{equation}
 \dfrac{\partial [(1-n)\rho_s]}{\partial t} + \nabla\cdot[(1-n)\rho_s\mathbf v_s] = 0
\label{solid_mass_balance}
\end{equation}
It now remains to derive the constitutive law for density of the solid phase.
This constitutive law can be derived considering the volumetric response of a porous element 
(in the following, equation 26a of \cite{Detournay_Cheng1993} is derived).
Consider a linear elastic porous medium of porosity $n$, saturated with fluid and loaded with an isotropic compressive
stress of $\Delta P_c$ under undrained conditions. This loading condition causes  within the specimen
mean total stress of magnitude
\begin{equation*}
\Delta\sigma = -\Delta P_c
\end{equation*}
(tensile stresses are positive), and increase in  pore pressure, $\Delta p$. 
The difference between the confining load and pore pressure is denoted with 
$\Delta P_d = \Delta P_c - \Delta p$.

Within the framework of an imaginary experiment, consider that the load is applied in two stages.
In the first stage  an increment of confining pressure 
$\Delta p$, and equal change in pore pressure  are applied (this is essentially the unjacketed test 
as described in Section \ref{sec:effective_stress}). Therefore at this stage $\Delta P_d = 0$.
In the second stage a confining load of $\Delta P_c - \Delta p$ is applied without any increase 
in the pore pressure, and $\Delta P_d = \Delta P_c - \Delta p$.

Consider the first stage of loading (where $\Delta P_d = 0$). It is convenient here to be reminded of the definition of the
compressibility of the solid phase, i.e. the unjacketed bulk compressibility
 
 \begin{equation}
   C_s = - \dfrac{1}{V} \dfrac{\Delta V}{\Delta p}|_{\Delta P_d = 0}
   \label{compress_solidPhase}
 \end{equation}
Another useful definition is that of the unjacketed pore compressibility
\begin{equation}
   C_{\phi} = - \dfrac{1}{Vp} \dfrac{\Delta Vp}{\Delta p}|_{\Delta P_d = 0}
   \label{compress_pore}
 \end{equation}
where $V_p = nV$ is the volume of pores. Compressibility $C_{\phi}$ is identified with $1/K_s''$ of \cite{rice1976}.
For saturated media this is equal to the volume of fluid phase.
Compressibility $ C_{\phi}$ is usually considered to be equal to $C_s$. However, this is true for porous materials of 
which the solid phase is composed of a single constituent - this requirement also excludes 
the presence of entrapped fluid and presence of cracks within the solid skeleton. For more details refer to \cite{wang2000}, Section 3.1.4.
Total volume is composed of the volume of the solid phase, $V_s = (1-n)V$, and that of the pores, $V_p = nV$, such that $V = V_c + V_p$.
The fractional volume change of the solid phase can be written as  

\begin{equation*}
   \dfrac{\Delta V_c}{V_c} = \dfrac{\Delta V}{V_c} - \dfrac{\Delta V_p}{V_c},
\end{equation*}
or
\begin{equation*}
   \dfrac{\Delta V_c}{V_c} = \dfrac{1}{1-n}\dfrac{\Delta V}{V} - \dfrac{n}{1-n}\dfrac{\Delta V_p}{V_p}
\end{equation*}
Definitions (\ref{compress_solidPhase}) and (\ref{compress_pore}) suggest substitutions $\dfrac{\Delta V}{V} = -C_s\Delta p$ and
$\dfrac{\Delta V_p}{V_p} = - C_{\phi}\Delta p$ into the above equation, yielding

\begin{equation}
   \dfrac{\Delta V_c}{V_c} = \dfrac{-C_s \Delta p + nC_{\phi}\Delta p }{1-n}
   \label{dvc_load1}
\end{equation}

In the second stage of loading the application of confining load leads to a mean effective stress increment in the solid phase equal to
$\Delta\sigma' = (\Delta P_c - \Delta p)/(1 - n)$. We can write $\Delta V_s/V_s = -C_s \Delta\sigma'$, which yields 

\begin{equation}
   \dfrac{\Delta V_c}{V_c} = \dfrac{-C_s (\Delta P_c - \Delta p) }{1-n}
   \label{dvc_load2}
\end{equation}

 Combining the effects in equations (\ref{dvc_load1}) and (\ref{dvc_load2}), and making the 
substitution $\Delta\sigma = -\Delta P_c$, the total volume change of the solid phase is obtained as

\begin{equation}
   \dfrac{\Delta V_c}{V_c} = \dfrac{C_s\Delta\sigma + nC_{\phi}\Delta p }{1-n}
   \label{dvc_total}
\end{equation}
(this is equation 26a of \cite{Detournay_Cheng1993}). Departing from equation (\ref{dvc_total}) and making the same 
arguments made for the derivation of the constitutive law for the fluid phase (\ref{fluid_compressibility_constitutive_1}),
the constitutive law for the solid phase can be derived as
\begin{equation}
   \dfrac{\partial \rho_s}{\partial t} = \dfrac{\rho_s}{1-n}\left(- C_s\dfrac{\partial\sigma}{\partial t} -nC_{\phi}\dfrac{\partial p}{\partial t}  \right)
\label{solid_density1}
\end{equation}
Substituting  
(\ref{solid_density1}) into (\ref{solid_mass_balance}) results to
\begin{equation}
 -\dfrac{\partial n}{\partial t} + \left( -C_s\dfrac{\partial\sigma}{\partial t} 
 -nC_{\phi}\dfrac{\partial p}{\partial t}  \right) + \nabla\cdot\mathbf v_s = \nabla\cdot(n\mathbf v_s)
\label{solid_mass_conservation}
\end{equation}

The term $\nabla\cdot(n\mathbf v_s)$ as calculated in equation (\ref{solid_mass_conservation}) is substituted into the equation 
of fluid mass conservation (\ref{fluid_mass_conservation}), that now reads

\begin{equation}
\nabla\cdot\mathbf v_s + \nabla\cdot\left[n(\mathbf v_f - \mathbf v_s) \right] + n(\beta - C_{\phi})\dfrac{\partial p}{\partial t} 
                               - C_s\dfrac{\partial\sigma}{\partial t} = 0
\label{storage1}
\end{equation}
The quantity $n(\mathbf v_f - \mathbf v_s)$ is the specific discharge defined in (\ref{discharge}),
and $\dfrac{\partial\epsilon}{\partial t}=\nabla\cdot\mathbf v_s$ is the time derivative of the volumetric strain.
Equation (\ref{storage1}) now is written as
\begin{equation}
\dfrac{\partial\epsilon}{\partial t}  + n(\beta - C_{\phi})\dfrac{\partial p}{\partial t} 
                   - C_s\dfrac{\partial\sigma}{\partial t} = -\nabla\cdot\mathbf q
\label{storage2}
\end{equation}
Equation (\ref{storage2}) can be further simplified by  substituting for the divergence of the specific 
discharge. Furthermore the mean (or isotropic) total stress $\sigma$ can be split into the mean effective stress 
and pore pressure components. The mean effective stress can be expressed with respect to the 
compressibility of the porous material  and the volumetric strain as 
\begin{equation}
 \sigma' = \dfrac{\epsilon}{C}
\end{equation}
Finally, making use of (\ref{alpha2}), fluid conservation equation is written as

\begin{equation}
\alpha\dfrac{\partial\epsilon}{\partial t}  + S_{\epsilon}\dfrac{\partial p}{\partial t} 
                    = -\nabla\cdot\mathbf q
\label{storage}
\end{equation}
In this form the equation of fluid mass conservation in fluid infiltrated porous media is known as the storage equation (\cite{verruijt1969}).
The term $-\nabla\cdot\mathbf q$ appearing in equation (\ref{storage}) is given by Darcy's law (\ref{darcy_0}), so that
equation (\ref{storage}) can be written as
\begin{equation}
\alpha\dfrac{\partial\epsilon}{\partial t}  + S_{\epsilon}\dfrac{\partial p}{\partial t} 
                    = \nabla\cdot\left(\dfrac{k}{\gamma_w}\nabla p \right)
\label{storage_pressure}
\end{equation}
The quantity $S_{\epsilon}$ appearing in the storage equation equals

\begin{equation}
  S_{\epsilon} =  \alpha C_s + n(\beta - C_{\phi})
\label{storativity_Cvoid}
\end{equation}
and, in literature, this is known as the storativity of the pore space. 
As will be seen in subsequent section, storativity can be expressed in terms of Skempton's pore pressure coefficient and Biot's coefficient 
(see equation \ref{storativity_2}). For practical purposes it is easier to calculate storativity by measuring Skempton's coefficient and
avoid the difficulties of measuring $C_{\phi}$.

For porous materials with solid phase composed of a single constituent, coefficients  $C_s$ and $C_{\phi}$ are equal
and $S_{\epsilon}$ is expressed as 
\begin{equation}
  S_{\epsilon} = n\beta + (\alpha - n)C_s
\label{storativity}
\end{equation}
Condition $C_s = C_{\phi}$ is commonly used in the literature.

\subsection*{Simplified expression of storage equation}
Storativity as expressed in equation (\ref{storativity}) depends on 
the compressibility of the pore fluid, $\beta$, and that  of the solid phase, $C_s$. In the consolidation problems of soil, 
the compressibility of the soil skeleton is considered to be zero, thus the storativity 
essentially depends only on the compressibility of the pore water. For saturated consolidating media containing only small
amounts of air bubbles, 
\cite{verruijt1969} proposed an upper bound  of the fluid compressibility as
\begin{equation}
 \beta = \dfrac{1}{K_f} + \dfrac{1-S_r}{p_w}
\label{fluid_compressibility}
\end{equation}
where $K_f \approx 2\cdot10^6 kN/m^2 $ is the water bulk modulus,  $S_r$ is the degree of saturation with liquid fluid,
and $p_w$ is the absolute pressure of the fluid, which can be taken as equal to 
the initial static pore pressure. The main assumption made in the derivation of (\ref{fluid_compressibility}) is that 
the degree of saturation should be close to unity, with $1 - S_r << 1$. 
The storage equation for soil mechanics problems can now be written as 
\begin{equation}
\dfrac{\partial\epsilon}{\partial t}  + n\beta\dfrac{\partial p}{\partial t} 
                    = \nabla\cdot\left(\dfrac{k}{\gamma_w}\nabla p \right)
\label{storage_soil}
\end{equation}
with $\beta$ defined in (\ref{fluid_compressibility}) for the case of fluid-gas mixtures and Biot's 
coefficient, $\alpha$, considered to be 1.


\section{Increment of fluid content}
\label{increment_of_fluid_content}

The equations of equilibrium and mass conservation were derived in Sections \ref{sec:equilibrium_eqns}  and 
\ref{mass_conservation} using pore pressure as a fundamental variable. The work conjugate variable to pore pressure
is the increment of fluid content, $\zeta$. It is defined as the change of fluid
volume per unit reference volume
\footnote{\cite{rice1976} defined the increment of fluid content as follows. Consider the mass of fluid, $M_f$, 
contained in a control volume, $V$, of porous material. Define the fluid mass content per unit volume as 
$m = M_f/V$, where $V$ is in an unstressed and unpressurised state. Define also the apparent fluid volume fraction as
$ \upsilon = m/\rho_o$, where $\rho_o$ is the reference state mass density of fluid. The increment of fluid content is simply 
\begin{equation*}
 \zeta = \upsilon - \upsilon_o
\end{equation*}
where $\upsilon_o$ is the reference value in the unstressed state.}.

In \cite{BiotWillis},  the increment of fluid content is quantified as
%
 %
 \begin{equation}
 \zeta = -n\nabla\cdot(\mathbf{U_f} - \mathbf{U_s})
\label{increment_fl_1}
\end{equation}
where  $\mathbf{U_f}$ and $\mathbf{U_s}$ signify the average displacement of the fluid and solid phases in the control volume.
Equation (\ref{increment_fl_1}) holds under the assumption that  porosity does not vary in space. Alternatively, $n$ should 
be put inside the parentheses in equation (\ref{increment_fl_1})  (\cite{wang2000}, Section $2.1.2$).
Taking the time derivative of the right hand side terms of (\ref{increment_fl_1}), results simply to the specific discharge velocity 
defined in equation (\ref{discharge}). Considering equations (\ref{increment_fl_1}) 
 and  (\ref{discharge}), in the absence of fluid sources, the following fluid continuity equation is found to hold 
\begin{equation}
 \dfrac{\partial \zeta}{\partial t} = -\nabla\cdot\mathbf{q} 
\label{increment_fl_time1}
\end{equation}
Next consider the storage equation (\ref{storage}), written in the form 
\begin{equation}
\dfrac{\partial}{\partial t}\left(\alpha\epsilon  + S_{\epsilon}p \right) 
                    = -\nabla\cdot\mathbf{q} 
\label{storage_n}
\end{equation}
From equations (\ref{storage}) and (\ref{increment_fl_time1}), and integrating with respect to time 
considering unstressed reference state, 
the constitutive relation for the increment of fluid content
is derived as 

\begin{equation}
 \zeta = \alpha\epsilon  + S_{\epsilon}p
\label{zeta_1}
\end{equation}
Introducing Biot's modulus, $M$, as the reciprocal of the storativity, $S_{\epsilon}$, such that 
\begin{equation}
 M = \dfrac{1}{S_{\epsilon}} 
\label{biot_modulus}
\end{equation}
the constitutive law  of the increment of fluid content (\ref{zeta_1}) takes the form
\begin{equation}
 \zeta = \alpha\epsilon  +\dfrac{1}{M}p
\label{zeta_2}
\end{equation}

\section{Fluid diffusion}
The equation of mass conservation can be expressed in terms of the increment of fluid content, $\zeta$, resulting 
in a diffusion type equation. Specifically, when the storage equation (\ref{increment_fl_time1}) 
is considered with Darcy's law (\ref{darcy_0}), and under the assumption
that $k/\gamma_w$ does not vary in space, it takes the form
\begin{equation}
 \dfrac{\partial \zeta}{\partial t} = \dfrac{k}{\gamma_w}\nabla^2p
\label{mass_zeta_1}
\end{equation}
Next,  the  right hand side of (\ref{mass_zeta_1}) is expressed in terms of  the increment of fluid content. 
The required expression can be derived from the equations of equilibrium (\ref{navier_index_1}). 
Taking the derivative of the $i-th$ equation of (\ref{navier_index_1}) with respect to $x_i$  and summing 
the resulting three equations, in the absence of body forces yields 

\begin{equation}
 \nabla^2\left(p - \dfrac{\lambda + 2G}{\alpha}\epsilon \right) = 0
 \label{stress_compatibility_zeta_1}
\end{equation}
Substituting equation (\ref{zeta_1}) into equation (\ref{stress_compatibility_zeta_1}) the required expression is obtained as
\begin{equation}
 \nabla^2 p = \dfrac{\lambda + 2G}{\alpha^2 + (\lambda + 2G)S_{\epsilon}} \nabla^2 \zeta 
 \label{stress_compatibility_zeta_3}
\end{equation}
Combining equations (\ref{mass_zeta_1}) and (\ref{stress_compatibility_zeta_3}), the equation of mass conservation 
takes the form of a diffusion equation as
\begin{equation}
 \dfrac{\partial \zeta}{\partial t} = c\nabla^2\zeta
\label{mass_diffusion}
\end{equation}
Equation (\ref{stress_compatibility_zeta_1}) used to derive the above equation was derived neglecting body forces. 
In the general case where body forces and fluid source are present, equation (\ref{mass_diffusion}) takes the form
\begin{equation}
 \dfrac{\partial \zeta}{\partial t} = c\nabla^2\zeta + \alpha m_v c\nabla\cdot\mathbf b + Q
\label{mass_diffusion_source}
\end{equation}
where $\mathbf b$ is the vector of body forces, coefficient $m_v$ is as defined in (\ref{volume_change}), and 
$Q$ is the fluid source term (for the mathematical treatment of the source term please refer to
\cite{wang2000}, \cite{Rudnicki1986}, or \cite{chau2012analytic}).
The coefficient, $c$,  is the coefficient of consolidation 
 in three dimensions, and is equal to
\begin{equation}
 c = \dfrac{k}{\gamma_w} \cdot \dfrac{\lambda + 2G}{\alpha^2 + (\lambda + 2G)S_{\epsilon}}
 \label{coefficient_of_consolidation_3d_1}
\end{equation}
or 
\begin{equation}
 c = \dfrac{k}{\gamma_w} \cdot \dfrac{(\lambda + 2G) M}{\alpha^2M + \lambda + 2G}
 \label{coefficient_of_consolidation_3d_1_2}
\end{equation}
where M is Biot's modulus defined in equation (\ref{biot_modulus}).

\section{Storage coefficients}

Two storage coefficients are examined in this section, storativity and the uniaxial storage coefficient.
Storativity, $S_\epsilon$, was introduced in equation (\ref{storage}) of mass conservation and its physical meaning is 
shown in this section. The uniaxial storage coefficient is the equivalent storage coefficient for laterally confined deformations
and is also introduced.

\subsection{ Physical interpretation of storativity}

Combining the equation of fluid mass conservation (\ref{storage}) with equation (\ref{increment_fl_time1}) yields
\begin{equation}
\alpha\dfrac{\partial\epsilon}{\partial t}  + S_{\epsilon}\dfrac{\partial p}{\partial t} 
                    = \dfrac{\partial \zeta}{\partial t}
\label{storage_def1}
\end{equation}
Pore pressure and volumetric strain depend on time, while the increment of fluid content
can be written as $\zeta(p(t), \epsilon(t))$. Its partial derivative with respect to time is
\begin{equation}
\dfrac{\partial \zeta}{\partial t} = \dfrac{\partial \zeta}{\partial p}\dfrac{\partial p}{\partial t} + 
                   \dfrac{\partial \zeta}{\partial \epsilon}\dfrac{\partial \epsilon}{\partial t}
\label{zeta_partial_t}
\end{equation}
For constant control volume the time derivative of volumetric strain vanishes from both equations (\ref{storage_def1})
and (\ref{zeta_partial_t}). Furthermore, substituting (\ref{zeta_partial_t}) to (\ref{storage_def1}) yields
\begin{equation}
\left(S_{\epsilon} - \dfrac{\partial \zeta}{\partial p}\right) \dfrac{\partial p}{\partial t} = 0
\label{zeta_partial_t2}
\end{equation}
The above equation involves multiplication of scalar quatities, from which the following relation is implied

\begin{equation}
S_{\epsilon} = \dfrac{\partial\zeta}{\partial p}\big\lvert_{\epsilon = 0}
\label{storage_def}
\end{equation}
Expression (\ref{storage_def}) gives to $S_{\epsilon}$ the inerpretation of being the fluid volume change per unit control volume and per unit pressure change, 
while the control volume remains constant.

\subsection{Uniaxial storage coefficient}
\label{sec:uniaxial_storage}

A storage coefficient for uniaxial deformation is defined in this section.
Depart from equation (\ref{storage_def1}) considering in addition that the control volume is confined laterally, i.e. 
$\epsilon_{xx} = 0$, and $\epsilon_{yy} = 0$. Equation (\ref{storage_def1}) then takes the form 

\begin{equation}
\alpha\dfrac{\partial\epsilon_{zz}}{\partial t}  + S_{\epsilon}\dfrac{\partial p}{\partial t} 
                    = \dfrac{\partial \zeta}{\partial t}
\label{storage_def2}
\end{equation}
Using the one-dimensional constitutive law ($\epsilon_{zz} = m_v\sigma'_{zz}$) and the principle of effective stress
($\sigma'_{zz} = \sigma_{zz} + \alpha p$), equation (\ref{storage_def2}) becomes

\begin{equation}
\alpha m_v \dfrac{\partial\sigma_{zz}}{\partial t}  + (S_{\epsilon} + \alpha^2 m_v)\dfrac{\partial p}{\partial t} 
                    = \dfrac{\partial \zeta}{\partial t}
\label{storage_def3}
\end{equation}
A new storage coefficient can now be defined as 
\begin{equation}
S =  S_{\epsilon} + \alpha^2 m_v
 \label{uniaxial_storage}
\end{equation}
Consider in addition that the total vertical stress $\sigma_{zz}$ is constant - then the term containing $\sigma_{zz}$
in equation (\ref{storage_def3}) vanishes. From (\ref{storage_def3}) and using arguments similar to the derivation of 
relation (\ref{storage_def}), $S$ can be defined as

\begin{equation}
S = \dfrac{\partial\zeta}{\partial p}\big\lvert_{\epsilon_{xx} = 0, \epsilon_{yy} = 0, \sigma_{zz} = c}
\label{uni_storage_def}
\end{equation}
The above expression gives $S$ the inerpretation of being the fluid volume change per unit control volume and per unit pressure change, 
while the control volume is confined in a state of zero lateral strain and constant vertical stress.

\subsubsection{Diffusion equation in one-dimensional consolidation}
\label{sec:uniaxial_cons}
This paragraph presents a generalisation of Terzaghi's 
consolidation. We make use of equations (\ref{storage_def3}) and (\ref{mass_zeta_1}). Substituting (\ref{mass_zeta_1})
into (\ref{storage_def3}) and rearranging yelds
\begin{equation}
  S\dfrac{\partial p}{\partial t} -\dfrac{k}{\gamma_w}\dfrac{\partial^2 p}{\partial z^2}
                                 = -\alpha m_v \dfrac{\partial\sigma_{zz}}{\partial t}
\label{gen_uniaxial_cons1}
\end{equation}
This equation of mass conservation is an inhomogeneous one dimensional equation of diffusion in 
terms of pore pressure, uncoupled from displacements. The right hand-side term is known
(total vertical stresses are equal to the externally applied vertical stresses),
and the pore pressure can be calculated independently from (\ref{gen_uniaxial_cons1}) without using 
the equations of equilibrium.
If the total vertical stress is kept constant, the right-hand side term of equation (\ref{gen_uniaxial_cons1})
vanishes. The result is Terzaghi's
diffusion equation
\begin{equation}
  \dfrac{\partial p}{\partial t} = \dfrac{k}{\gamma_w S}\dfrac{\partial^2 p}{\partial z^2}
 \label{terzaghi_diffusion}
\end{equation}
where the coefficient of consolidation acounts for material compressibility and is given by 

\begin{equation}
 c = \dfrac{k}{\gamma_w S} 
 \label{coefficient_of_consolidation_3d}
\end{equation}
Based on  equation (\ref{coefficient_of_consolidation_3d_1}) another expression is found for 
the uniaxial storage coefficient, that is
\begin{equation}
 \dfrac{1}{S} =  \dfrac{\lambda + 2G}{\alpha^2 + (\lambda + 2G)S_{\epsilon}}
 \label{uniaxial_storage_2}
\end{equation}

\section{Undrained description of poroelastic equations}

In the undrained description of poroelastic equations 
the increment of fluid content is introduced as a fundamental variable, associated with undrained poroelastic
constants.  For this description the equation of fluid mass balance has 
already been derived (see equation \ref{mass_diffusion}). Stress-strain relations,  equilibrium equations, and 
undrained poroelastic constants are introduced in the remainder of this section.


\subsection{Skempton's pore pressure coefficient B}
\label{sec:coeff_B}
%
%
%
The pore pressure coefficient,  B, is used to find the increase in  pore 
 pressure developed in the pores of an elastic isotropic porous material under undrained loading
with an all round confining pressure.
Assume that an isotropic linear elastic porous medium filled with fluid is loaded with isotropic compressive
stress $\Delta P_c$ under undrained conditions ($\zeta = 0$). This loading  causes mean total stress of 
magnitute $\Delta\sigma = -\Delta P_c$ (tensile stresses are positive), and increase in  pore 
pressure, $\Delta p$. The increase in pore pressure under this type 
of loading is related to the isotropic stress and coefficient B as
\begin{equation}
 B = -\dfrac{\Delta p}{\Delta \sigma}|_{\zeta = 0}
 \label{skempton_coeff_1}
\end{equation}
The use of pore pressure coefficient, B, as a material property was first proposed by \cite{skemptonB}.
A general analytical expression can be derived as follows. Substituting equation (\ref{mean_eff_stress_2}) in quation (\ref{zeta_1}) 
the constitutive relation for the increment of fluid content becomes

\begin{equation*}
  \zeta = \dfrac{\alpha}{K}\sigma + \dfrac{\alpha^2 +KS_{\epsilon}}{K}p
\end{equation*}
Making use of definition (\ref{skempton_coeff_1}), the above equation yields 
\begin{equation}
 B = \dfrac{\alpha}{\alpha^2 + KS_{\epsilon}}
 \label{Skempton_coeff_2}
\end{equation}
%
%
Using expression (\ref{storativity_Cvoid}) for the storage coefficient $S_{\epsilon}$, equation (\ref{Skempton_coeff_2}) 
can be written as 
%
%
\begin{equation}
 B = \dfrac{C - C_s}{C - C_s + n(\beta -C_{\phi})}
 \label{Bishop_B_0}
\end{equation}
In a more familiar form used in the field of soil mechanics, pore pressure coefficient can be obtained substituting
expression (\ref{storativity}) into equation (\ref{Skempton_coeff_2}) as

\begin{equation}
 B = \dfrac{C - C_s}{C - C_s + n(\beta -C_{s})}
 \label{Bishop_B}
\end{equation}
In this form pore pressure coefficient  was first derived by \cite{bishop1973}.

\subsection*{Storativity in terms of the pore pressure coefficient B}

Storativity can be 
expressed as an equation of B and of the drained bulk modulus (or compressibility) of the porous medium. 
From equation ( \ref{Skempton_coeff_2}) the following expression is obtained
\begin{equation}
  S_{\epsilon} = \dfrac{\alpha }{B}\dfrac{(1-\alpha B)}{K}
\label{storativity_2}
\end{equation}
The term $\frac{K}{1-\alpha B}$ will be defined in equation (\ref{k_undrained}) 
as the undrained bulk modulus of the porous medium, $K_u$, which makes equation (\ref{storativity_2}) equal to
\begin{equation}
  S_{\epsilon} = \dfrac{\alpha }{BK_u}
\label{storativity_3}
\end{equation}


\subsection{Undrained poroelastic constants}
\label{undrained_parameters}

Consider undrained conditions, i.e. there is no change in the fluid content of the control volume ($\zeta = 0$).
We can define the undrained bulk modulus such that 
\begin{equation}
 \sigma = K_u\epsilon
\label{k_undrained_1}
\end{equation}
with $\sigma = \dfrac{\sigma_{kk}}{3}$.The stress-strain relation for a saturated medium was derived in 
equation (\ref{stress_strain_total_3d}). We also make use of relation 
\footnote{Which is derived making use of  equations  (\ref{zeta_1}) and (\ref{storativity_2}).}
\begin{equation}
 \zeta = \dfrac{\alpha}{K}\sigma  +\dfrac{a}{KB}p
 \label{zeta_B_K}
\end{equation}
and substitute  for $\alpha p$ to (\ref{stress_strain_total_3d}). For undrained conditions, where 
$\zeta = 0$, this results to

\begin{equation}
\sigma_{ij}  = 2G\epsilon_{ij}  + \left(K - \dfrac{2G}{3}\right)\epsilon\delta_{ij}  + \alpha B \sigma \delta_{ij}
\label{stress_strain_undrained_def_1}
\end{equation}
 We can derive the values of shear and bulk moduli 
from the above system of equations and the definition in (\ref{k_undrained_1}).
First, it is easy to see from (\ref{stress_strain_undrained_def_1}) that 
\begin{eqnarray*}
  \sigma_{ij}  =  2G\epsilon_{ij},   & & \textrm{for i $\neq$ j}\
\end{eqnarray*}
which is the same with the case of drained deformation. This means that shear modulus is the same in drained and undrained 
deformations. 

The undrained bulk modulus can be derived as follows. From (\ref{stress_strain_undrained_def_1}) the contracted volumetric 
constitutive equation is
\begin{equation}
   \sigma =  \dfrac{K}{1 - \alpha B} \epsilon
\label{k_undrained_0}
\end{equation}
Invoking the definition in (\ref{k_undrained_1}) we can conclude that the undrained bulk modulus can be expressed as
\begin{equation}
   K_u =  \dfrac{K}{1 - \alpha B} 
\label{k_undrained}
\end{equation}
Using the above equation and equation (\ref{storativity_3}) a new relation for Biot's modulus can be derived as
\begin{equation}
   M =  \dfrac{B^2K_u^2}{K_u - K} 
\label{biot_M_ku}
\end{equation}

The rest of undrained elasticity constants can be defined similar to classical elasticity theory. 
The undrained Lam\'{e} constant is defined as

\begin{equation}
  \lambda_u = K_u - \dfrac{2}{3}G
\label{lame_undrained}
\end{equation}
and it is  directly equivalent to the equation for the drained Lam\'{e} constant
\begin{equation}
  \lambda = K - \dfrac{2}{3}G
\label{lame_drained}
\end{equation}
Subtracting equation (\ref{lame_drained}) from (\ref{lame_undrained}) provides
\begin{equation}
 \lambda_u - \lambda = K_u - K
\label{Lame_undrained_1}
\end{equation}
which, with the aid of equation (\ref{k_undrained})   results to the expression  
\begin{equation}
 \lambda_u  = \lambda + \alpha B K_u 
\label{Lame_undrained_2}
\end{equation}
Another useful expression for the undrained Lam\'{e} constant can be derived from equation (\ref{Lame_undrained_1}) using equations 
(\ref{storativity_3})  and (\ref{biot_modulus}) and reads
\begin{equation}
 \lambda_u  = \lambda + \alpha^2 M
\label{Lame_undrained_3}
\end{equation}
Similarly we can derive the following relation for the undrained bulk modulus, known as Gassman's equation

\begin{equation}
 K_u  = K + \alpha^2 M = K + \dfrac{\alpha^2}{aC_s + n(\beta - C_{\phi})}
\label{k_undrained_gassman}
\end{equation}
The undrained Poisson's ratio is defined as  
\begin{equation}
  \nu_u = \dfrac{\lambda_u} {2(\lambda_u + G)}
\label{poisson_undrained_4}
\end{equation}
A useful expression for the undrained Poisson ratio 
can be derived using equations (\ref{poisson_undrained_4}), (\ref{Lame_undrained_2}), (\ref{k_undrained}), 
and the third of equations (\ref{shear_modulus_2}). It expresses $\nu_u$ in terms of the pore pressure 
coefficient B  as
\begin{equation}
  \nu_u = \dfrac{3\nu + \alpha B(1-2\nu)}{3 - \alpha B(1-2\nu)}
\label{poisson_undrained}
\end{equation}
Equation (\ref{poisson_undrained}) was first derived by \cite{rice1976}.
Lastly, departing from the relations $G_u = G$ and $G = \dfrac{E}{2(1+\nu)}$, the undrained Young modulus is defined as 
\begin{equation}
  E_u = \dfrac{1+\nu_u}{1+\nu}E
 \label{undrained_E}
\end{equation}
Further useful relations between undrained constants are the following 
\begin{equation}
   K_u = \dfrac{E_u}{3(1-2\nu_u)}; \hspace{0.5cm} \lambda_u = \dfrac{E_u\nu_u}{(1+\nu_u)(1-2\nu_u)}; 
   \hspace{0.5cm} K_u = G\dfrac{2(1+\nu_u)}{3(1-2\nu_u)}
\label{undrained_shear_modulus}
\end{equation}

\subsection{Constitutive relations}
\label{undrained_const_law}

We are now ready to derive the undrained desctiption of the constitutive relations 
(in terms of the increment of fluid content and undrained elastic constants). The stress-strain relations
are derived making use of equations (\ref{stress_strain_total_3d}) 
in which pressure is substituted from equation (\ref{zeta_2}), yielding

\begin{equation}
\sigma_{ij}  = 2G\epsilon_{ij}  + \lambda_u\epsilon\delta_{ij}  - \alpha M \zeta \delta_{ij}
\label{stress_strain_undrained_descr_1}
\end{equation}
Furthermore, strain-stress relations are derived from (\ref{strain_stress_total_3d}) using equations (\ref{zeta_2})
and (\ref{poisson_undrained}) as

\begin{equation}
\epsilon_{ij}  =  \dfrac{1}{2G} \left(\sigma_{ij} - \dfrac{\nu_u}{1 + \nu_u}\sigma_{kk}\delta_{ij} \right) 
              + \dfrac{B}{3} \zeta\delta_{ij} 
\label{strain_stress_total_3d_1}
\end{equation}

\subsection{Pore pressure coefficient for uniaxial strain}

\begin{figure}
	\centering
		\includegraphics[width=0.65\textwidth]{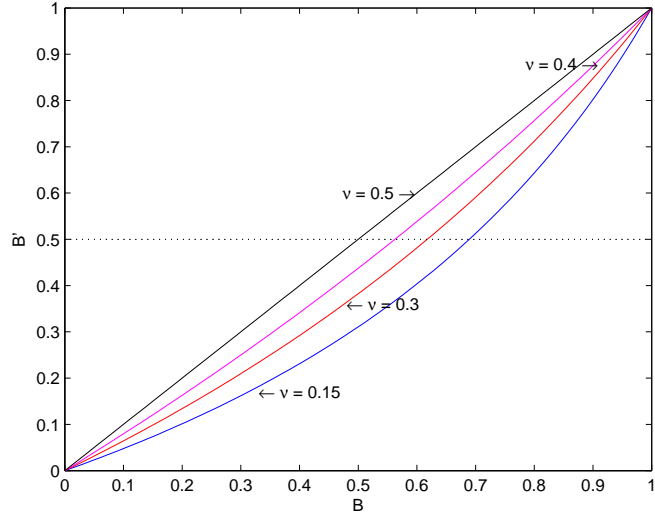}
	\caption{Relationship between Skempton's coefficient B, and the uniaxial pore pressure coefficient, $B'$.}
	\label{fig:B_vs_Buniaxial}
\end{figure}

A pore pressure coefficient, $B'$, equivalent to Skempton's coefficient, $B$, can be 
defined for the case of uniaxial strain load under undrained conditions (\cite{lancellotta2008geotechnical}, 
equation (6.51)). This coefficient is also known as loading efficiency (\cite{wang2000}, Section 3.6.2).
Suppose that the porous medium is restricted in the $x$ and $y$ directions  and an external load, $\delta\sigma_{zz}$,
is applied. In this case, the increase of the pore pressure, $\delta p$, under undrained conditions is 
\begin{equation}
 B' = -\dfrac{\delta p}{\delta\sigma_{zz}}\mid_{\epsilon_{xx} =\epsilon_{yy} = \zeta = 0}
 \label{B_uniaxial_definition}
\end{equation}
The negative sign in equation (\ref{B_uniaxial_definition}) is a consequence of the definition of tensile stresses as positive.
A positive increase in the uniaxial stress causes a negative change in the pore pressure which is positive when it is compressive.
An expression for $B'$  can be derived as follows
\footnote{A similar approach to the one adopted here (which leads to equation (\ref{B_uniaxial})) can be found 
in \cite{lancellotta2008geotechnical}.  Lancelotta considers that the fluid flux is zero
 ($-\nabla \mathbf q =0$) and he integrates the  equation of mass conservation for zero initial conditions.}.
The undrained response of a porous medium implies that $\zeta = 0$.
$\zeta$ is given by the constitutive law 
(\ref{zeta_1}) and, since for the case of uniaxial strain the volumetric strain equals
$\epsilon = \epsilon_{zz}$, equation (\ref{zeta_1}) is transformed into
\begin{equation}
  \alpha\epsilon_{zz}  + S_{\epsilon}p = 0
\label{zeta_undrained_2}
\end{equation}
Using the one-dimensional constitutive law ($\epsilon_{zz} = m_v\sigma'_{zz}$) and the principle of effective stress
($\sigma'_{zz} = \sigma_{zz} + \alpha p$), equation (\ref{zeta_undrained_2}) leads to the reformulation
\begin{equation}
  m_v\sigma_{zz}  = \left(\alpha m_v +\dfrac{ S_{\epsilon}}{\alpha}\right)p 
\label{zeta_undrained_3}
\end{equation}
From equations (\ref{B_uniaxial_definition}) and (\ref{zeta_undrained_3}) can be concluded that the uniaxial pore pressure 
coefficient is equal to
\begin{equation}
 B' = \dfrac{m_v}{\alpha m_v + \dfrac{S_{\epsilon}}{\alpha}}
 \label{B_uniaxial}
\end{equation}
$B'$ can be expressed in terms of  Skempton's coefficient $B$ by substituting for $m_v$ and $S_{\epsilon}$ into
equation (\ref{B_uniaxial}). To obtain this result,  equation (\ref{B_uniaxial}) is written in the form 
\begin{equation}
 \dfrac{1}{B'} =  \dfrac{S_{\epsilon}m_v}{\alpha} + \alpha
 \label{B_uniaxial_1}
\end{equation}
The coefficient of one-dimensional compressibility can be expressed in terms of the bulk modulus and Poisson's ratio
 as
\begin{equation}
 m_v = \dfrac{1+\nu}{3(1-\nu)K}
 \label{volume_change2}
\end{equation}
Substituting equations (\ref{volume_change2}) and (\ref{storativity_2}) into equation (\ref{B_uniaxial_1})
results to
\footnote{In the case of $\alpha = 1$, equation (\ref{B_uniaxial_2}) can be found in  \cite{okusa85}.} 
\begin{equation}
 \dfrac{1}{B'} =  \dfrac{1}{B}+\dfrac{2(1-2\nu)(1-\alpha B)}{B(1+\nu)}
 \label{B_uniaxial_2}
\end{equation}
Inverting equation  (\ref{B_uniaxial_2}) the coefficient $B'$ reads
\begin{equation}
   B' =  \dfrac{B(1+\nu)}{(1+\nu) +2(1-2\nu)(1-\alpha B)}
 \label{B_uniaxial_3}
\end{equation}
The variation of $B'$ with respect to $B$ for various values of Poisson's ratio is plotted in figure \ref{fig:B_vs_Buniaxial}.
For Poisson's ratio equal to $\nu=1/2$, the two coefficients coincide; for other values of Poisson's ratio,
$B'$ is smaller than $B$.

There is a second approach to defining the  uniaxial pore pressure coefficient $B'$, which 
can be found  in Section 3.6.2 of \cite{wang2000}.
Making use of the stress-strain relation in terms of the increment of fluid content (equation \ref{stress_strain_undrained_descr_1}),
for undrained behaviour (where $\zeta = 0$) and
under uniaxial strain, the conditions $\epsilon_{xx} = 0$ and $\epsilon_{yy} = 0$ yield
\begin{equation}
  \sigma_{xx} = \sigma_{yy} = \dfrac{\nu_u}{1-\nu_u}\sigma_{zz}
\label{strain_stress_undrained_x}
\end{equation}
Equations  (\ref{strain_stress_undrained_x})  lead to the mean stress being expressed as
\begin{equation}
  \sigma_{kk} = \dfrac{1+\nu_u}{1-\nu_u}\sigma_{zz}
\label{B_uniaxial_undrained_1}
\end{equation}
From the definition of the pore pressure coefficient, B, provided in equation (\ref{skempton_coeff_1}), the pore pressure
applied to the porous medium under conditions of uniaxial strain is
\begin{equation}
  3p = -\dfrac{B(1+\nu_u)}{1-\nu_u}\sigma_{zz}
\label{B_uniaxial_undrained_2}
\end{equation}
Using the definition (\ref{B_uniaxial_definition}), the coefficient $B'$ can now be calculated from 
(\ref{B_uniaxial_undrained_1}) as being
\begin{equation}
  B' = \dfrac{1}{3}\dfrac{B(1+\nu_u)}{(1-\nu_u)}
\label{B_uniaxial_undrained}
\end{equation} 
For incompressible media, the undrained Poisson ratio turns out to be $\nu_u = 0.5$ 
(see equation \ref{poisson_undrained_2}) and $B' = B$. For highly compressible material,
$\nu_u$ approaches zero, and $B'\approx B/3$.

At this point it is interesting to turn the attention once again on the coefficient of consolidation.
The coefficient of  consolidation was expressed in equation (\ref{coefficient_of_consolidation_3d_1}) in terms of
 Lam\'{e}'s constants $\lambda$ and $G$ and the storativity $S_{\epsilon}$.  
In view of the equation (\ref{volume_change1}) of the one dimensional compressibility, the coefficient of consolidation
can be written with respect to $m_v$ as
\begin{equation}
 c = \dfrac{k}{\gamma_w}\cdot \dfrac{1}{m_v\left(\alpha^2 +\dfrac{S_{\epsilon}}{m_v}\right)}
 \label{coefficient_of_consolidation_3d_2}
\end{equation}
Using equation (\ref{B_uniaxial}), equation (\ref{coefficient_of_consolidation_3d_2}) can be further 
modified to take the form
\begin{equation}
 c = \dfrac{k}{\gamma_w  m_v} \dfrac{B'}{\alpha}
 \label{coefficient_of_consolidation_3d_3}
\end{equation}
Defining the poroelastic stress coefficient 
\begin{equation}
  \eta = \dfrac{\alpha(1-2\nu)}{2(1-\nu)}
 \label{por_stress_coeff}
\end{equation}
(\cite{wang2000}), then
\begin{equation*}
 \alpha m_v = \eta/G
\end{equation*}
Using the above equation an additional expression of the coefficient of consolidation is 
derived as
\begin{equation}
 c = \dfrac{k}{\gamma_w }\dfrac{G}{\eta} B'
 \label{coefficient_of_consolidation_3d_4}
\end{equation}
Finally, comparing equations (\ref{coefficient_of_consolidation_3d_4}) and (\ref{coefficient_of_consolidation_3d}), a new expression
for the uniaxial pore pressure coefficient can be derived, and reads
\begin{equation}
 B' = \dfrac{\eta}{GS}
 \label{B_uniaxial2}
\end{equation}

\subsection{Undrained Poisson ratio}
\label{sec:undrained_parameters}

Equation (\ref{poisson_undrained_4}) offers an expression for the undrained Poisson ratio.   
A second important expression for the undrained Poisson ratio  links it to the pore pressure coefficient, $B$, and 
Poisson's ratio, $\nu$, as recorded in equation (\ref{poisson_undrained}). For its derivation we can also use equations 
(\ref{B_uniaxial_undrained}) and 
(\ref{B_uniaxial_3}), equating their right hand sides and performing some algebraic manipulations,
the expression for the undrained Poisson ratio reads

\begin{equation*}
  \nu_u = \dfrac{3\nu + \alpha B(1-2\nu)}{3 - \alpha B(1-2\nu)}
\label{poisson_undrained_ksana}
\end{equation*}
Another useful expression of the undrained Poisson ratio can be obtained in terms of  storativity $S_{\epsilon}$
by using equations (\ref{poisson_undrained}) and (\ref{storativity_2}). Equation (\ref{storativity_2})
can be rewritten in the form
\begin{equation}
 B = \dfrac{\alpha}{KS_{\epsilon} + \alpha^2}
\label{B_2}
\end{equation}
which, when substituted into equation (\ref{poisson_undrained}) yields

\begin{equation}
  \nu_u = \dfrac{\lambda S_{\epsilon} + \alpha^2} {2(\lambda + G)S_{\epsilon} + 2\alpha^2}
\label{poisson_undrained_2}
\end{equation}
where $K$ has been substituted for $\lambda + 2G/3$. Lastly, using Biot's modulus instead of storativity the above equation becomes
or 
\begin{equation}
  \nu_u = \dfrac{\lambda + \alpha^2 M} {2(\lambda + G)+ 2\alpha^2M}
\label{poisson_undrained_3}
\end{equation}

\subsection{Equations of equilibrium}
\label{equilibrium_zeta}

Equilibrium in terms of displacements and  pore pressure was examined in Section $2.4.4$.
The resulting Navier equations (\ref{navier_index_1}) are
\begin{equation}
  G \nabla^2 u_i + \left(\lambda + G\right) \dfrac{\partial\epsilon}{\partial x_i}
                 = \alpha\dfrac{\partial p}{\partial x_i}  -b_i
\label{navier_index_2}
\end{equation}
Equilibrium equations  (\ref{navier_index_2}) use  pore pressure  as a fundamental variable.
Instead of pore pressure, the increment of fluid content can be used  as a fundamental variable, 
leading to an undrained description of the equilibrium. Using equation (\ref{zeta_2}), the pressure 
can be expressed in terms of the increment of fluid content as
\begin{equation}
 p = M\zeta - \alpha M \epsilon
\label{zeta_press}
\end{equation}
Substituting equation (\ref{zeta_press}) into (\ref{navier_index_2}) yields
\begin{equation}
  G \nabla^2 u_i + \left(\lambda + \alpha^2M + G\right) \dfrac{\partial\epsilon}{\partial x_i}
                 = \alpha M\dfrac{\partial \zeta}{\partial x_i}  -b_i
\label{navier_index_3}
\end{equation}
which, due to the relation (\ref{Lame_undrained_3}) can be written as
\begin{equation}
  G \nabla^2 u_i + \left(\lambda_u + G\right) \dfrac{\partial\epsilon}{\partial x_i}
                 = \alpha M\dfrac{\partial \zeta}{\partial x_i}  -b_i
\label{navier_index_4}
\end{equation}
Since the relation $\lambda_u + G = G/(1-2\nu_u)$ holds, equilibrium equations can also appear as
\begin{equation}
  G \nabla^2 u_i + \dfrac{G}{1-2\nu_u}\dfrac{\partial\epsilon}{\partial x_i}
                 = \alpha M\dfrac{\partial \zeta}{\partial x_i}  -b_i
\label{navier_index_5}
\end{equation}

\subsection*{Mean stress formulation of the undrained description of equilibrium}
\label{mean_stress_zeta}

The formulation of equilibrium (\ref{navier_index_5}) in terms of $\zeta$ and $\nu_u$, 
is not valid for $\nu_u = 1/2$ (i.e. for both pore fluid and soil particles being incompressible)
\footnote{Provided there is fluid flow (case in which always $\nu < 1/2$), 
                 equations (\ref{navier_index_2}) can be used instead.}. 
In the present subsection   a mean stress formulation of the equilibrium equations in terms of $\zeta$ and $\nu_u$
is derived. Although 
the specific formulation is complicated to use for the formulation of general consolidation problems, in this work it
proves particularly useful in the consolidation of porous media with incompressible constituents. Specifically, use of 
this formulation is made in \cite{Merxhani_Thesis}, where  the completeness 
of displacement functions appropriate for consolidation problems is proven.
The pair of constitutive equations  (\ref{mean_eff_stress_2}) - (\ref{stress_strain_mean_stress}) is used
\begin{equation}
\sigma_{ij}  =  2G\epsilon_{ij} + \dfrac{3\nu}{1+\nu}\sigma\delta_{ij} - \dfrac{2G\alpha }{3K}p \delta_{ij} 
							      \hspace{0.2cm} \text{and} \hspace{0.5cm} \sigma + \alpha p = K\epsilon
\label{stress_strain_mean_stress_1}
\end{equation}
This is similar to the pair of constitutive equations 
(\ref{stress_strain_const_incompress1})  for incompressible elasticity.
Mean stress, $\sigma$, appearing in them is treated as an independent variable which, however, can always
be eliminated by substituting the right hand side equation of (\ref{stress_strain_mean_stress_1}) into the left hand side
system of equations, resulting to
the stress-strain relations (\ref{stress_strain_eff_3d}).
From the above pair of constitutive equations the following one can be derived.
When equation (\ref{zeta_B_K}) is substituted for $\alpha p$ 
\begin{equation}
\sigma_{ij}  =  2G\epsilon_{ij} + \dfrac{3\nu_u}{1+\nu_u}\sigma\delta_{ij} - \dfrac{2GB}{3}\zeta \delta_{ij}
\label{stress_strain_mean_stress_3}
\end{equation}
and
\begin{equation}
    \epsilon  = \dfrac{3}{2G}\dfrac{(1-2\nu_u)}{(1+\nu_u)}\sigma + B\zeta
 \label{mean_1}
\end{equation}
 (\ref{stress_strain_mean_stress_3}) is derived from (\ref{stress_strain_mean_stress_1}) making  use 
of equations (\ref{zeta_B_K}), (\ref{Lame_undrained_2}),  (\ref{Lame_undrained_1}), and  (\ref{lame_undrained}). 
Equation (\ref{mean_1}) can be calculated  directly from (\ref{stress_strain_mean_stress_3}).
Given the constitutive relation  (\ref{stress_strain_mean_stress_3}), the equations of equilibrium can be derived
using the same process with Section \ref{sec:incompressible_elasticity}.
In the absence of body forces, the equations of equilibrium in terms of stresses are  
 \begin{equation}
 \dfrac{\partial\sigma_{ij}}{\partial x_j} = 0
 \label{mean_stress_eq__zeta_1}
\end{equation}
The stress-strain relation (\ref{stress_strain_mean_stress_3}) can be used with the equilibrium
 equations  (\ref{mean_stress_eq__zeta_1}) to yield
\begin{equation}
   2G\dfrac{\partial \epsilon_{ij} }{\partial x_j} + \dfrac{3\nu_u}{1+\nu_u}\dfrac{\partial\sigma}{\partial x_i}  = 
                                                        \dfrac{2GB}{3}\dfrac{\partial \zeta }{\partial x_i}    
 \label{eq_mean_zeta1}
\end{equation}
or
\begin{equation}
   G  \dfrac{\partial^2 u_i}{\partial x_j \partial x_j} +    
  G\dfrac{\partial}{\partial x_i}\left(\dfrac{\partial u_j} {\partial x_j}\right)  
   + \dfrac{3\nu_u}{1+\nu_u}\dfrac{\partial\sigma}{\partial x_i}  = \dfrac{2GB}{3}\dfrac{\partial \zeta }{\partial x_i}    
 \label{eq_mean_zeta2}
\end{equation}
Since
\begin{equation}
 \nabla^2(\cdot) = \dfrac{\partial^2(\cdot)}{\partial x_j \partial x_j} ; \hspace{0.5cm} \epsilon = \dfrac{\partial u_j} {\partial x_j}
\end{equation}
the equations of equilibrium  read
\begin{equation}
   G  \nabla^2u_i + G\dfrac{\partial}{\partial x_i}\epsilon 
   + \dfrac{3\nu_u}{1+\nu_u}\dfrac{\partial\sigma}{\partial x_i}  =  \dfrac{2GB}{3}\dfrac{\partial \zeta}{\partial x_i}
 \label{eq_mean_zeta3}
\end{equation}
with
\begin{equation}
    \epsilon  = \dfrac{3}{2G}\dfrac{(1-2\nu_u)}{(1+\nu_u)}\sigma + B\zeta
 \label{mean_2}
\end{equation}
For an incompressible constituent model, where $S_{\epsilon} = 0$ and $\alpha = 1$, the undrained Poisson ratio
is equal to $\nu_u = 1/2$, and the pore pressure coefficient is  $B = 1$. 
Therefore, for incompressible material behaviour, equilibrium is given by the system
\begin{equation}
 G  \nabla^2u_i + \dfrac{\partial\sigma}{\partial x_i}  = - \dfrac{1}{3}G\dfrac{\partial \zeta}{\partial x_i}, 
                  \hspace{0.5cm}  \epsilon = \zeta
 \label{eq_mean_zeta_inc}
\end{equation}

\newpage

\section{Variational formulation and FEM}
\label{sec:weak_form}

The main numerical method used to solve  consolidation problems is the finite element method. 
In this method, an appropriate discretisation is applied to the weak formulation of 
the problem. A weak formulation for consolidation problems is derived in this section, and is associated 
with the finite element discretisation of the problem. The weak formulation is derived for drained
description of equations with primary variables displacements and pore pressure ($\mathbf{u} - p$ formulation).

\subsection{Strong form}
The strong form of a problem of isothermal consolidation consists of the equations of equilibrium and the 
equation of conservation of mass, associated with appropriate boundary conditions.
In the present formulation, the equations of equilibrium as defined in Section \ref{sec:equilibrium_eqns} are used,
while the storage equation  (\ref{storage}) is used to express conservation of mass. Therefore, the primary variables of the 
problem are the displacements $\mathbf{u}$ and the pore pressure p.

First, define a domain $\Omega$ and the boundary of the domain $\Gamma$.  Next, define the portions of 
the boundary $\Gamma_u$ 
and $\Gamma_t$ on which displacements and stresses are defined, such as $\Gamma_u \cup \Gamma_t = \Gamma$ and
\begin{equation}
\hspace{0.2cm} \mathbf{u} = \bar{\mathbf{u}}  \hspace{0.2cm} on \hspace{0.2cm} \Gamma_u, \hspace{0.2cm}  and \hspace{0.2cm} 
               \mathbf{t} = \bar{\mathbf{t}},   \hspace{0.2cm} on \hspace{0.2cm} \Gamma_t 
\label{u_boundary}
\end{equation}
The portions of the boundary $\Gamma_p$ and $\Gamma_q$ are the parts of the boundary in which pressure and pressure flux 
are specified, with $\Gamma_p \cup \Gamma_q = \Gamma$ and
\begin{equation}
  p = \bar{p} \hspace{0.2cm} on \hspace{0.2cm} \Gamma_p, \hspace{0.2cm} and 
                                  \hspace{0.2cm} \dfrac{\partial p}{\partial n} = \bar q  \hspace{0.2cm} on \hspace{0.2cm} \Gamma_q
\label{p_boundary}
\end{equation} 
The traction boundary condition on (\ref{u_boundary}) is defined such as $t_i = \sigma_{ij}n_i$, with $\mathbf{n}$ being the
outer  unit vector perpendicular to the surface $\Gamma$. Using index notation, the components of the surface 
traction are given by the relations
\begin{equation}
 t_i = (\sigma'_{ij} - \alpha \delta_{ij}p)n_i
\label{traction}
\end{equation}
Expanding the components of eq.(\ref{traction}) for the two dimensional case, the following relations are obtained
\begin{equation}
\begin{array}{l}
 t_x = \sigma_{xx}n_x +  \sigma_{xz}n_z =  (\sigma'_{xx} - ap)n_x +  \sigma_{xz}n_z \\
 
 t_z = \sigma_{zx}n_x +  \sigma_{zz}n_z =    \sigma_{zx}n_x + (\sigma'_{zz} - ap)n_z
\end{array}
\label{traction_2d}
\end{equation}
From (\ref{traction}) is concluded that both effective stresses and pore pressure should be defined as a loading boundary 
condition for the displacements.

The consolidation problem for porous linear elastic materials is defined by equations (\ref{equilibrium_matrix_form1}), 
(\ref{effective_stress_matrix}), (\ref{constitutive1}), and (\ref{storage})
\begin{equation}
 \begin{array}{l}
   \nabla^T_s\mathbf{\sigma} + \mathbf{b }= 0 \\
  \mathbf{\sigma} = \mathbf{\sigma'} - \alpha\mathbf{m} p, \hspace{0.2cm}
   \mathbf{\sigma'} = D\nabla_s\mathbf{u} \\
 \end{array}
\label{consolidation_strong}
\end{equation}
\begin{equation}
 \begin{array}{l}
      \alpha\dfrac{\partial\epsilon}{\partial t}  + S_{\epsilon}\dfrac{\partial p}{\partial t}  
                    = \nabla\cdot\left(\dfrac{k}{\gamma_w}\nabla p \right)
 \end{array}
\label{consolidation_strong_2}
\end{equation}
and the boundary conditions
\begin{equation}
 \begin{array}{l}
    \mathbf u = \bar{\mathbf u}  \hspace{0.2cm} on \hspace{0.2cm} \Gamma_u, \hspace{0.2cm} 
    and \hspace{0.2cm} \mathbf t = \bar{\mathbf t} \hspace{0.2cm} on \hspace{0.2cm} \Gamma_t \\
    
     p = \bar{p} \hspace{0.2cm} on \hspace{0.2cm} \Gamma_p,\hspace{0.2cm} 
     and \hspace{0.2cm} \dfrac{\partial p}{\partial n} = \bar q  \hspace{0.2cm} on \hspace{0.2cm} \Gamma_q
     
 \end{array}
\label{consolidation_boundary}
\end{equation}
Equations $(\ref{consolidation_strong})$ lead to the equilibrium equations (\ref{equilibrium_final_3d})
or (\ref{navier_index_1}), for the three-dimensional or the plane strain problem respectively, 
and are associated with the boundary conditions $(\ref{consolidation_boundary}_{,1})$, while the boundary 
conditions $(\ref{consolidation_boundary}_{,2})$ are associated with the storage equation 
$(\ref{consolidation_strong_2})$.

\subsection{Weak form }
The weak form is derived separately for the equilibrium equations and  the storage equation.
In order to derive the weak form of the equilibrium equations, it is more convenient to express equation  
$(\ref{consolidation_strong}_{,1})$ 
in index notation. Multiplying this equation by an arbitrary function, $\delta \mathbf u$, such that
 $\delta \mathbf u = 0 \hspace{0.2cm} $on$ \hspace{0.2cm} \Gamma_u$ and integrating over the domain results in
\begin{equation}
  \int\limits_{\Omega} \delta u_i \sigma_{ij,j} \text{ }d\Omega + \int\limits_{\Omega} \delta u_i b_i \text{ }d\Omega = 0
\label{weak_eq_1}
\end{equation}
Integrating by parts the first term of (\ref{weak_eq_1}) leads to
\begin{equation}
  -\int\limits_{\Omega} \delta u_{i,j} \sigma_{ij} \text{ }d\Omega + \int\limits_{\Omega} (\delta u_i \sigma_{ij})_{,j} \text{ }d\Omega
   + \int\limits_{\Omega} \delta u_i b_i \text{ }d\Omega = 0
\label{weak_eq_2}
\end{equation}
Splitting the tensor $\delta u_{i,j}$ into symmetric and antisymmetric parts, the multiplication of its antisymmetric 
part  by $\sigma_{ij}$ - which
is symmetric - results in zero. Furthermore using the Gauss divergence theorem and the condition that 
$\delta  u_i = 0 \hspace{0.2cm} $on$ \hspace{0.2cm} \Gamma_u$, equation (\ref{weak_eq_2}) becomes
\begin{equation}
  -\int\limits_{\Omega} \dfrac{1}{2}(\delta u_{i,j} + \delta u_{j,i}) \sigma_{ij} \text{ }d\Omega 
             + \int\limits_{\Gamma_t} \delta u_i \sigma_{ij}n_j \text{ }d\Gamma
   + \int\limits_{\Omega} \delta u_i b_i \text{ }d\Omega = 0
\label{weak_eq_3}
\end{equation}
Equation (\ref{weak_eq_3}) is written in  matrix form as 
\begin{equation}
  \int\limits_{\Omega} (\nabla_s\delta \mathbf u)^T\mathbf \sigma \text{ }d\Omega = 
      \int\limits_{\Gamma_t} (\delta \mathbf{u})^T \bar{\mathbf{t}}\text{ }d\Gamma
   + \int\limits_{\Omega} (\delta \mathbf u)^T \mathbf b \text{ }d\Omega 
\label{weak_eq_4}
\end{equation}
where the symmetric gradient operator $\nabla_s$ was introduced in Section \ref{sec:equilibrium_eqns}, 
and the boundary traction, $\bar{\mathbf t}$, was defined in equations (\ref{traction}) and 
$(\ref{consolidation_boundary}_{,2})$. Finally, using equations $(\ref{consolidation_strong}_{,2})$,
 the weak form of the equilibrium equation is obtained:
\begin{equation}
  \int\limits_{\Omega} (\nabla_s\delta \mathbf u)^TD\nabla_s \mathbf u  \text{ }d\Omega 
    -  \int\limits_{\Omega} (\nabla_s\delta \mathbf u)^Ta\mathbf m p  \text{ }d\Omega
                       = \int\limits_{\Gamma_t} (\delta \mathbf{u})^T \bar{\mathbf{t}}\text{ }d\Gamma
                          + \int\limits_{\Omega} (\delta \mathbf u)^T \mathbf b \text{ }d\Omega
\label{weak_equilibrium}
\end{equation}
Following \cite{zienkiewicz2005finite}, the solid skeleton displacements, $\mathbf u$, the weighting function, 
$\delta \mathbf u$, and the pore fluid pressure, $p$,  are discretised as 
\begin{equation}
 \mathbf u = \mathbf N \tilde{\mathbf u}, \hspace{0.5cm} \delta \mathbf u = \mathbf N\delta \tilde{\mathbf u},  
 \hspace{0.5cm} p = \mathbf N_p \tilde{\mathbf p}
\label{dicretise_1}
\end{equation}
with $\mathbf N$ and $\mathbf N_p$ being the shape functions of $\mathbf u$ and $p$
and $\tilde{\mathbf u}$ and $\tilde{\mathbf p}$ representing the values at the element
nodes. The arbitrariness of $\delta \mathbf u$ implies that $\delta \tilde{\mathbf u}$ 
is arbitrary as well. Substituting (\ref{dicretise_1}) into (\ref{weak_equilibrium}) yields
\begin{equation}
\delta \tilde{\mathbf u} \left( \mathbf K\tilde{\mathbf u} - \mathbf Q\tilde{\mathbf p} - \mathbf f \right) = 0
  \label{weak_eq_dicretisation1}
\end{equation}
with the arbitrariness of $\delta \tilde{\mathbf u}$ implying that 
\begin{equation}
 \mathbf K\tilde{\mathbf u} - \mathbf Q\tilde{\mathbf p} - \mathbf f  = 0
\label{weak_dicretisation1}
\end{equation}
The matrices $\mathbf K$ and $\mathbf Q$ and the vector, $\mathbf f$, appearing in equation (\ref{weak_dicretisation1}) 
are specified as
\begin{equation}
\begin{array}{l}
  \mathbf K = \int\limits_{\Omega} (\nabla_s\mathbf N)^TD\nabla_s \mathbf N\text{ }d\Omega, \hspace{0.4cm} 
  \mathbf Q = \int\limits_{\Omega} (\nabla_s\mathbf N)^Ta\mathbf m \mathbf{N_p}  \text{ }d\Omega \\
   
  and \hspace{0.9cm} \mathbf f = \int\limits_{\Gamma_t} \mathbf N^T \bar{\mathbf{t}}\text{ }d\Gamma
                          + \int\limits_{\Omega} \mathbf N^T  \mathbf b \text{ }d\Omega
\end{array}
\label{weak_form_components_1}
\end{equation}
The weak form of the storage equation is derived next.
The storage equation (\ref{storage}) can be written as 
\begin{equation}
 - \nabla\cdot\left(\dfrac{k}{\gamma_w}\nabla p \right) + \alpha\dot\epsilon  + S_{\epsilon}\dot p = 0
\label{weak_storage_1}
\end{equation}
 with the upper dot denoting derivation in time. The time derivative of the volumetric strain, $\dot\epsilon$, can be expressed as 
\begin{equation}
  \dot\epsilon = \mathbf m^T \dot{\mathbf \epsilon} = \mathbf m^T \nabla_s\dot{\mathbf u} 
\label{volumetric}
\end{equation}
with $\mathbf \epsilon$, $\mathbf m$,  $\mathbf u$, and $\mathbf \nabla_s$ defined in Section \ref{sec:equilibrium_eqns}.
 Multiplying (\ref{weak_storage_1}) by an arbitrary function,  $\delta p$, such that 
 $\delta p = 0 \hspace{0.2cm} $on$ \hspace{0.2cm} \Gamma_p$, and integrating over the domain results in
\begin{equation}
 -  \int\limits_{\Omega}(\delta p)^T \nabla\cdot\left(\dfrac{k}{\gamma_w}\nabla p \right)d\Omega  
+ \int\limits_{\Omega}(\delta p)^T\alpha\mathbf m^T \nabla_s\dot{\mathbf u}d\Omega 
+  \int\limits_{\Omega}(\delta p)^T S_{\epsilon}\dot pd\Omega = 0
\label{weak_storage_2}
\end{equation}
Integrating by parts the first term of (\ref{weak_storage_2}) and applying the divergence theorem results in the weak form of the 
storage equation
\begin{equation}
   \int\limits_{\Omega}(\delta p)^T\alpha\mathbf m^T \nabla_s\dot{\mathbf u}d\Omega   
+  \int\limits_{\Omega}(\delta p)^T S_{\epsilon}\dot pd\Omega
+   \int\limits_{\Omega}(\nabla\delta p)^T \dfrac{k}{\gamma_w}\nabla p d\Omega  
 =   \int\limits_{\Gamma_q}(\delta p)^T \dfrac{k}{\gamma_w}\bar q d\Gamma
\label{weak_storage}
\end{equation}
where the condition $\delta p = 0$ on $\Gamma_p$ is used.
The parameter $\bar q$ on the right hand side term of (\ref{weak_storage}) is the pore pressure flux on the boundary $\Gamma_q$,
defined in equation $(\ref{consolidation_boundary}_{,2})$. 
Using the same discretisation technique as in equation (\ref{weak_equilibrium}),
equation (\ref{weak_storage}) results in the discretised scheme
\begin{equation}
 \mathbf Q^T\dot{\tilde{\mathbf u}} + \mathbf S\dot{\tilde{\mathbf p}} + \mathbf H\tilde{\mathbf p} - \mathbf q  = 0
\label{weak_dicretisation2}
\end{equation}
where $\mathbf Q$ was defined in equation (\ref{weak_form_components_1}). The rest of the terms that appear in equation 
(\ref{weak_dicretisation2}) are defined as
\begin{equation}
\begin{array}{l}
  \mathbf S =  \int\limits_{\Omega}(\nabla \mathbf N_p)^T S_{\epsilon}\nabla \mathbf N_p d\Omega , \hspace{0.4cm} 
  \mathbf H = \int\limits_{\Omega}\mathbf N^T_p \dfrac{k}{\gamma_w}\nabla \mathbf N_p d\Omega  \\
   
  and \hspace{0.9cm} \mathbf q = \int\limits_{\Gamma_q}\mathbf N^T_p \dfrac{k}{\gamma_w}\bar q d\Gamma
\end{array}
\label{weak_form_components_2}
\end{equation}
The discrete coupled system of equations (\ref{weak_dicretisation1}) and (\ref{weak_dicretisation2}) can now be written as
\begin{equation}
 \begin{bmatrix}  \mathbf 0 & \mathbf 0 \\  \mathbf Q^T & \mathbf S \end{bmatrix} 
 \begin{bmatrix}  \dot{\tilde{\mathbf u}} \\  \dot{\tilde{\mathbf p}} \end{bmatrix} + 
\begin{bmatrix}  \mathbf K & -\mathbf Q \\  \mathbf 0 & \mathbf H \end{bmatrix}  \begin{bmatrix}  \tilde{\mathbf u} \\ 
\tilde{\mathbf p} \end{bmatrix} = 
\begin{bmatrix}  \mathbf f \\  \mathbf q \end{bmatrix}
\label{discretised_weak_form}
\end{equation}

\subsection{Restrictions on the choice of discretisation basis}
\label{subsec:Taylor-Hood}

The storativity, $S_{\epsilon}$, that was defined in (\ref{storativity}), often approaches zero in consolidation problems. 
This is due to the incompressibility of the solid constituent, which is always the case in soil mechanics problems
and, due to the near incompressibility of the pore fluid, in the absence of air. When these conditions hold, then
\begin{equation}
   \mathbf S \approx \mathbf 0
\label{undrained_1}
\end{equation}
Furthermore, if the permeability is small enough such that
\begin{equation}
   \mathbf H \approx \mathbf 0
\label{undrained_2}
\end{equation}
the behaviour of the consolidated system is practically undrained and the problem defined by the system 
(\ref{discretised_weak_form}) 
is  of a saddle point type (\cite{zienkiewiczBook,zienkiewicz2005finite}). 
Approximating the variables of a saddle point problem  with the same shape functions  results in the loss 
of coercivity of the numerical approximation, and the error of the approximation is no longer bounded 
(\cite{ern2004theory}). The problem becomes stable when the Babuska-Brezzi condition 
(\cite{Brezzi74}), or the equivalent patch test of \cite{zienkiewiczPatchTest} is satisfied. 
The satisfaction of the Babuska-Brezzi condition requires  the  pressure field to be approximated with 
polynomials of a lower degree than the polynomials approximating  the displacement field. This is 
the reason why different shape functions $\mathbf N$ and $\mathbf N_p$ were used in (\ref{dicretise_1}).  
A compatible pair of polynomial approximation is that of a quadratic approximation of displacements and 
a linear approximation of the pressure. An example of an element satisfying the Babuska-Brezzi condition 
is the Taylor-Hood element, which is widely used in modelling 
problems of saddle point type, for example like the ones of incompressible fluid flow.

\newpage

\section{Epilogue}

The theory of linear poroelasticity is derived for isotropic fluid saturated media.     
The material considered is isotropic,  and undergoes quasi-static deformations under 
isothermal conditions. Porous space is considered to be connected - however, the existence 
of isolated voids or cracks within the solid skeleton is not excluded. Solid phase is 
compressible and is not necessarily composed of a single constituent. Furthermore, pore 
fluid is compressible and consists of a single phase. 

Both drained and undrained descriptions of poroelasticity are presented. The exposition 
starts with Verruijt's drained formulation of poroelasticity - this approach uses a direct proof of 
equations of equilibrium and fluid mass balance using the principle of effective stress
and Darcy's law of fluid flow in porous media.
So far pore pressure is used as an independent variable in the description. The
only new coefficients that are necessarily introduced in the equations are Biot’s
coefficient that appears in the principle of effective stress, the storage coefficient
that appears in the equation of fluid mass balance, and definitions of material compressibility. 
In the derivation of fluid mass balance presented, storativity 
(and consequently Skempton's pore pressure coefficient) is expressed in a more general form than in
Verruijt's original formulation, since it also includes the effect of the unjacketed 
pore compressibility.
The undrained description of poroelasticity is derived leading to Rice and Cleary's 
formalism. The constitutive law for the increment of fluid content as derived follows 
naturally from the equation of fluid mass balance. 
Pore pressure coefficients, $B$ and $B'$, and uniaxial storage coefficient, $S$, are 
derived from the constitutive equation of the increment of fluid content. Other 
undrained poroelastic coefficients are also introduced ($K_u, \nu_u \lambda_u, E_u$) 
and useful relations are proven. Where appropriate, the physical meaning of these 
coefficients is proven mathematically. 
Lastly, a weak formulation is derived for drained description of equations with primary 
variables displacements and pore pressure ($\mathbf u - p$ formulation).

Texts that can be consulted to further readers' knowledge of poroelasticity include the following.
\cite{verruijt2008, verruijt_2013}, who approaches poroelasticity using a soil mechanics viewpoint,
and the general reviews of \cite{Detournay_Cheng1993} and \cite{chau2012analytic}. 
The book of \cite{wang2000} that provides a complete text on the subject with applications 
to geomechanics and hydrogeology. Furthermore, \cite{coussy2004poromechanics} offers an advanced treatment
of poromechanics, where the theory is systematically derived departing from a general formulation in 
terms of large deformations. In the above citations a plethora of examples is also included.

\bibliographystyle{abbrvnat}
\bibliography{references}

\end{document}